# Coherent excitation of heterosymmetric spin waves with ultrashort wavelengths

G. Dieterle[1,*], J. Förster[1], H. Stoll[1], A. S. Semisalova[2], S. Finizio[3], A. Gangwar[4], M. Weigand[1], M. Noske[1], M. Fähnle[1], I. Bykova[1], J. Gräfe[1], D. A. Bozhko[5], H. Yu. Musiienko-Shmarova[5], V. Tiberkevich[6], A. N. Slavin[6], C. H. Back[4], J. Raabe[3], G. Schütz[1] and S. Wintz[2,3,*]

[1] Max-Planck-Institut für Intelligente Systeme, 70569 Stuttgart, Germany
[2] Helmholtz-Zentrum Dresden-Rossendorf, 01328 Dresden, Germany
[3] Paul Scherrer Institut, 5232 Villigen PSI, Switzerland
[4] Universität Regensburg, 93053 Regensburg, Germany
[5] Technische Universität Kaiserslautern, 67663 Kaiserslautern, Germany
[6] Oakland University, Rochester, MI 48309, USA
[*] e-mail: dieterle@is.mpg.de; sebastian.wintz@psi.ch

**In the emerging field of magnonics, spin waves are foreseen as signal carriers for future spintronic information processing and communication devices, owing to both the very low power losses and a high device miniaturization potential predicted for short-wavelength spin waves. Yet, the efficient excitation and controlled propagation of nanoscale spin waves remains a severe challenge. Here, we report the observation of high-amplitude, ultrashort dipole-exchange spin waves (down to 80 nm wavelength at 10 GHz frequency) in a ferromagnetic single layer system, coherently excited by the driven dynamics of a spin vortex core. We used time-resolved x-ray microscopy to directly image such propagating spin waves and their excitation over a wide range of frequencies. By further analysis, we found that these waves exhibit a heterosymmetric mode profile, involving regions with anti-Larmor precession sense and purely linear magnetic oscillation. In particular, this mode profile consists of dynamic vortices with laterally alternating helicity, leading to a partial magnetic flux closure over the film thickness, which is explained by a strong and unexpected mode hybridization. This spin-wave phenomenon observed is a general effect inherent to the dynamics of sufficiently thick ferromagnetic single layer films, independent of the specific excitation method employed.**

Spin waves are the collective excitations of magnetically ordered spin systems, with the 'magnon' being their fundamental quantum of excitation [1-3]. In a simplified view, a spin-wave can be regarded as a collective precession of the magnetization ($M$) with a periodic spatial phase shift [Fig. 1a], imposing the wavelength $\lambda$ and, inversely, the wavevector $k = (2\pi/\lambda)e_k$, with $e_k$ corresponding to the propagation direction of phase fronts. Taking into account the precession frequency $f$, the phase velocity of a spin-wave is determined by $v = \lambda f$. Depending on the underlying magnetic system, spin waves cover a wide spectral band, up to the THz range with wavelengths extending from macroscopic length scales to the sub-nanometer domain [4,5]. While spin waves also occur in three-dimensional bulk systems, the main focus of magnonics research is set on quasi two-dimensional and technologically most relevant thin film systems.

The two main advantages of using spin waves over present charge-based technologies are a substantially lower power consumption due to the absence of ohmic losses in the signal transfer, and a high device miniaturization potential owing to the orders of magnitude shorter wavelengths of spin waves compared to that of electromagnetic waves [4-14]. Prior to further implementation, however, the coherent excitation and controlled propagation of nanoscale spin waves remains a challenge to be resolved.

Typical methods to excite spin waves coherently in magnetic thin films utilize non-linear parametric pumping by uniformly alternating magnetic fields [9,14], patterned conductive antennas as microwave to spin-wave transducers [7,8,14], or current induced spin-transfer torques in nano-contact geometries [9,10,12]. A major drawback of parametric pumping, however, is the cumbersome tunability of the excited wavelengths, whereas the two other methods mentioned experience difficulties to efficiently generate propagating spin waves with wavelengths below the patterning sizes involved [14]. Recently, the latter limitation was partially overcome in specific magnetic systems by -when categorized in general terms- making use of Fano resonances [6,11,15-18], exploiting the Schlömann mechanism [19-23], or by a post-excitation variation of the magnonic index [24-26]. Yet, for the technologically most relevant plain thin film geometry, the demand for high amplitude ultrashort spin-wave excitation sources and a subsequent wave propagation remains unanswered at present.

In general, the dispersion relation of spin waves, i.e. the functional relation between frequency and wavelength, is mainly determined by both the magneto-dipole and the exchange interaction, where the first one is dominant in the long wavelength limit and the second governs the regime of ultrashort waves [14,27-29]. In magnetic thin films, the dispersion relation strongly depends on the relative orientation of the propagation direction and the static equilibrium magnetization. For in-plane magnetized films, there is a distinction between the so-called backward-volume ($k \parallel M$) and the Damon-Eshbach ($k \perp M$) geometry, where the latter typically exhibits higher phase velocities and positive group velocities [14,30], making it preferable for signal transmission applications.

While in principle being of surface wave nature with an exponentially decaying thickness profile in bulk samples [30], Damon-Eshbach waves exhibit an almost uniform distribution of the wave amplitude over the film thickness in thin films whose thicknesses $d$ are much smaller than $\lambda$, as shown in Fig. 1**b**. Such waves have been investigated by optical means, namely by magneto-optical Kerr microscopy [31] and Brillouin light scattering [32], so far down to wavelengths at the optically accessible lower limit of $\lambda \sim 250$ nm. In addition to this quasi-uniform mode, exchange dominated *higher order* spin waves with non-uniform mode profiles over the film thickness occur at higher frequencies due to vertical confinement effects. Such modes are typically observed in ferromagnetic resonance experiments as perpendicular standing waves for the non-propagating case of laterally uniform precession $(\lambda_{xy} \to \infty)$ [33,34], and they are characterized by their number of precession nodes along the film thickness, as shown in Fig. 1**c** (one node) [27,28]. Nevertheless, also first experimental evidences have been found for laterally propagating *higher order* waves of finite $\lambda$ in the Damon-Eshbach geometry [27-29,35,36], by means of spectroscopically detecting incoherent magnons of the thermal excitation spectrum [37-39] or even coherent waves yet only in the long wavelength limit [40-42].

In this Article, we show that high-amplitude, ultrashort ($\lambda \sim 100$ nm) propagating spin waves can be excited in a single layer magnetic thin film with a thickness of the order of 100 nm by a nanoscopic magnetic vortex core. For employing this excitation process, which was pioneered recently in a multi-layered heterosystem [6], the vortex core is driven to gyration by applying alternating external magnetic fields with frequencies of the order of $f \sim 10$ GHz, and the resulting spin-wave wavelength was found to be directly tuneable by the driving frequency. We used time-resolved magnetic x-ray microscopy to image the effects of spin-wave emission and propagation. By further analysis, we show that the observed spin waves correspond to the first *higher order* mode (having one thickness node) in the Damon-Eshbach geometry. Interestingly, the thickness profile of the observed mode is found to be influenced by a significant mode hybridization and that it therefore is of heterosymmetric character with respect to the perpendicular (symmetric) and lateral (antisymmetric) dynamic magnetization components. Thus, the excited spin wave mode exhibits points of purely linear magnetic oscillations and regions with anti-Larmor precession sense.

Our experimental sample is a Permalloy ($Ni_{81}Fe_{19}$) circular thin film disc with a lateral diameter of 3 μm and a thickness of $d = 80$ nm [cf. supplemental material (SM) (1)]. The expected magnetic ground state for this sample is that of a topological spin vortex with flux-closing in-plane magnetization circulation and a perpendicularly oriented nanoscopic vortex core in the center [43]. We verified the sample to be in such a magnetic vortex state by imaging its local magnetic orientation ***m*** = ***M***/*M* using scanning transmission x-ray microscopy (cf. methods). Figure 2**a** shows a microscopy image with lateral $m_x$-sensitivity (white: $+m_x$, black: $-m_x$) revealing the in-plane magnetic orientation in the disc (indicated by the red curling arrow).

For the generation of spin waves, the sample was excited by in-plane magnetic fields $H_y(t) = H_{y0} \sin(2\pi f t)$ alternating in time *t*. We directly imaged the dynamic response ***m****(t)* in the disc by stroboscopic, time-resolved scanning transmission x-ray microscopy in the frequency range from 5.6 to 10.1 GHz. Figure 2**b** displays a snapshot of such a dynamic response with perpendicular $m_z$ magnetic sensitivity to an excitation with a frequency of $f = 7.4$ GHz and an amplitude of $\mu_0 H_{y0} \sim 1$ mT, ($\mu_0 = 4\pi \cdot 10^{-7}$ Vs/Am being the vacuum permeability). The vortex core is clearly visible as a black dot in the center of the disc (pointing into the plane). Moreover, a radial spin-wave pattern can be observed, with highest amplitudes in the vicinity of the core (cf. magnified inset), decaying towards the edge of the disc. This spin-wave pattern is further highlighted in a normalized view (cf. Fig. 2**c**), showing the temporal perpendicular magnetic deviations [$\Delta m_z(t)$] in the disc. From the full dynamic microscopy image sets, provided as movies (M1-M6) in the SM, it becomes obvious that the observed spin waves are emitted from the vortex core. Driven by the alternating magnetic field, the gyrating vortex core dynamically induces dips of same and opposite magnetic orientation acting as gyrating pair of confined perpendicular magnetic perturbations [44-48]. These perturbations cause the coherent local excitation of propagating waves [6] [cf. SM (3d)] resulting in a spiralling emission pattern that was further confirmed by micromagnetic simulations [cf. SM M7]. More abstractly, this effect can be seen as a local Fano resonance [15] of the discrete core gyration mode with the continuum of propagating spin waves, mediated via a coherent and linear coupling. Note that such linear coherent spin-wave generation is not a consequence of a frequency-doubling mechanism [49] and, at the same time, fundamentally different from the process of incoherent spin-wave excitation reported earlier for the case of dynamic vortex core reversal [50,51]. Once excited, the spin waves in our experiment propagate radially towards the edge of the disc, i.e., the propagation is always perpendicular to the azimuthal vortex equilibrium magnetization, which formally corresponds to the Damon-Eshbach geometry ($\mathbf{k} \perp \mathbf{M}$). Along with the continuous phase difference of the spin waves emitted during gyration, this leads to the formation of a spiral pattern. The propagation is visualized by means of line profiles (along the green arrow in Fig. 2**c**) for a relative time delay of 77 ps as indicated by a blue and a red line, respectively, in Figure 2**d**. From this graph we conclude that the spin-wave amplitude exhibits very high values in the vicinity of the core [$m_z \approx 0.25$, cf. Fig 2**d**, SM M1] and that the wavelength is about 140 nm, leading to a phase velocity of 1040 m/s.

Moreover, we found that the wavelengths of such excited spin waves can be continuously tuned by changing the driving frequency in the range from 5.6 GHz to 10.1 GHz without the need of applying any external magnetic bias field, as shown in Figures 2**e** and 2**f** exemplarily for ($f$ = 5.6 GHz, $\lambda$ = 250 nm) and ($f$ = 8.5 GHz, $\lambda$ = 110 nm), respectively. This wide-band process of vortex core driven spin-wave generation, observed in a single magnetic layer, underlines the universality of the effect of spin wave emission from confined non-collinear spin textures such as domain walls [49,52-55] and topological vortex cores, where the latter had only been reported for a highly specific multi-layered heterosystem so far [6]. Besides the fact that the spin wave excitation takes place in a single magnetic layer, the excitable spin-wave frequencies found here (exceeding 10 GHz), are substantially higher than those observed in the previous case [6], and the corresponding wavelengths clearly attain the ultrashort (sub-100 nm) regime. Note also that the observed spin waves are not standing wave eigenmodes of the disc [46], but are independent from the lateral sample dimensions [cf. SM (3c)].

The spin-wave dispersion relation $f(k)$ of the experimentally observed spin waves [cf. Fig. 3, red dots] was found to exhibit an almost linear behavior at a constant group velocity of $v_g = d\omega/dk$ = 550 m/s in the $k$-range given (25<$k$<80) rad/µm, where $\omega = 2\pi f$ corresponds to the angular frequency. Remarkably, this dispersion is qualitatively different from the analytically calculated zeroth order, quasi-uniform (for k < 25 rad/µm) Damon-Eshbach dispersion [27,28] [cf. SM(4)], (see blue line in Figure 3) typically investigated in the framework of magnonics so far.

At the same time, *higher order* perpendicular standing spin-wave modes observed for the case of laterally uniform precession $(\lambda_{xy} \rightarrow \infty)$ [33,34] are predicted to extend to the regime of laterally propagating dipole-exchange waves of finite $\lambda_{xy}$. We therefore analytically calculated [27] (cf. methods) the dispersion relations of both the first ($p$=1) and the second ($p$=2) of these *higher order* modes, shown as red and green lines in Fig. 3, respectively, with $p$ being the ordinal index of nodes in the mode thickness profile at zero lateral wave number $k_{xy}$ = 0 [27]. This given, a clear matching of $f_1(k)$ with the experimental results can be seen. We therefore identify the experimentally observed spin wave mode as the first *higher order propagating mode*. The experimental observation of coherent, *higher order propagating* spin waves is further supported by an excellent agreement of both experimental and analytical results with micromagnetic simulations [blue triangles: quasi-uniform mode, red crosses: first *higher order* mode in Fig. 3, SM (5)]. At the same time, the simulations show that, in the given (*f,k*)-range, the dispersion relations for plane spin waves in a continuous thin film and for radial waves in a magnetic vortex state are practically identical [cf. SM (5)], which underlines the general validity of our findings. In addition to first higher order waves, also quasi-uniform waves with longer wavelengths (λ>200 nm) can in principle be excited in the disc structures used [cf. SM(5)]. For that case, however, the waves are not excited by vortex core gyration but stem from the magnetic discontinuity at the rim of the disc.

In general, hybridization (or mixing) between modes having different ordinal indices *p* has little effect on their dispersion relations up to moderate *k*-values. However, hybridization might become relevant for higher *k* values and it is crucially important in the vicinity of mode crossing points, eventually leading to avoided crossings when hybridization is taken into account analytically (cf. inset Fig. 3 quasi-uniform/*higher order* mode) [27]. Yet neglecting hybridization, the dispersion relations $f_p(k)$ of the dipole-exchange modes of higher order (*p≥1*) can be analytically calculated, yielding for the case of free surface boundary conditions [27, SM (4a)]:

$$f_p(k) = \frac{\gamma\mu_0}{2\pi}\sqrt{\left(\frac{2A}{\mu_0 M_S}\left(k^2 + \left(\frac{p\pi}{d}\right)^2\right)\right)\left(\frac{2A}{\mu_0 M_S}\left(k^2 + \left(\frac{p\pi}{d}\right)^2\right) + M_S\right)} \qquad (1)$$

with $\gamma = 176.8$ GHz/T being the gyromagnetic ratio and $A = 0.75\cdot 10^{-11}$ J/m being the exchange constant [cf. SM(4)]. Note that dispersion relations $f_p(k)$ strongly depend on both *A* and *d*, i.e., the dispersion relation can be efficiently tuned by varying these parameters [cf. SM (4)].

It is remarkable that for sufficiently thick layers, as it is the case for our system, the *first higher order* spin wave dispersion above a certain frequency (here ~5 GHz) exhibits significantly larger *k*-values than the quasi-uniform Damon-Eshbach dispersion (cf. Fig. 3). For *k*-values above ~ 1 rad/um, equivalently, the frequency (and thus the magnon energy) of the *first higher order* spin wave mode is lower than that of the *quasi-uniform* one. This is in contrast to ferromagnetic resonance experiments $(\lambda_{xy} \to \infty)$ and travelling waves in ultrathin films, where *higher order* perpendicular standing waves *solely* occur at frequencies above that of the uniform precession and the quasi-uniform propagation, respectively [33,34].

In order to qualitatively understand the characteristics of the experimentally observed dispersion law, we evaluated the simulated thickness profiles for both the quasi-uniform (cf. Fig. 4 left column **a-c**) and the first *higher order* spin wave mode (cf. Fig. 4 right column, **d-f**) at a fixed frequency of 8 GHz (dotted horizontal line in Fig. 3). With the static equilibrium magnetization $m_{eq}$ aligned along +$y$ (into the paper plane), the top (middle) panels of Fig. 4 show the thickness distribution of the dynamic magnetization components $m_z$ ($m_x$) at a fixed time, while the bottom panels display the corresponding precession orbits and phases. By analyzing the quasi-uniform spin-wave profile with a lateral wavelength of ~ 2 μm, the following becomes obvious: As expected, both dynamic magnetization components $m_z$ (Fig. 4a) and $m_x$ (Fig. 4b) are distributed almost uniformly over the film thickness, with a slight decay from the top to the bottom surface only. As a consequence, the precession phase (Fig. 4c) is constant over the film thickness, and the sense of precession always corresponds to the Larmor sense, i.e. a right-handed precession with respect to the direction of the equilibrium magnetization $m_{eq}$.

This situation fundamentally changes for the first *higher order* mode, which exhibits a more than 20 times shorter wavelength of only ~ 100 nm at the same frequency. Here, the dynamic magnetic $m_z$ component appears to be much more localized in the center of the film, decaying towards both the top and bottom surfaces (cf. Fig. 4d). The position of the maximum amplitude, nevertheless, is slightly shifted from the center to the lower half-space of the film. Analogously to quasi-uniform Damon-Eshbach waves, but with opposite sign, this amplitude asymmetry determines the wave propagation direction in relation to $m_{eq}$, which means $k$ is pointing along +$x$ for the given case. The dynamic $m_x$-component profile (cf. Fig. 4e) of the *higher order* mode, however, is found to exhibit a nodal line of zero $m_x$-amplitude in the upper half-space of the film, close to its center. Above and below this nodal line, the corresponding $m_x$ components exhibit an antiphase relation while their amplitudes increase towards the respective surfaces of the film. In the same panel, arrows indicate the local $m_{x,z}$ components, revealing that the dynamic magnetization is following a flux-closing, thus stray-field reducing, vortex over the film thickness. Laterally these vortices alternate in their circulation sense (one pair per wavelength) and their propagating cores are showing zero dynamic components with $m = m_y$. This heterosymmetric mode profile (0 $m_z$-nodes, 1 $m_x$-node) has crucial implications for the spin-wave energy and the local magnetization precession shown in Fig. 4f. While in the film region above the $m_x$-nodal line a mildly elliptic precession with Larmor sense can be observed, the dynamics on the node line itself corresponds to a purely linear magnetic oscillation along the $z$ axis. Most strikingly, however, a precession with reversed Larmor sense is exhibited in the film region above the nodal line. Although such local anti-Larmor precession was theoretically predicted already in the 1970s [36] and thereafter discussed in the context of incoherent thermal magnon measurements [35], it had not been experimentally verified for the case of short-wavelength coherent spin waves so far. Note that such behavior is also qualitatively different from the *higher order* modes both observed in ferromagnetic resonance experiments and analytically calculated neglecting mode hybridization, for which the precession always has Larmor sense and only the phase is inverted along the thickness profile.

From a more formal perspective, the dynamic magnetization orientation $\widetilde{m}_{xz,p,k}(x,z,t)$ of a plane spin-wave with wave vector $k$ and the order index $p$ in the Damon-Eshbach geometry ($\boldsymbol{m}_{eq} = \boldsymbol{m}_y, \boldsymbol{k} = \boldsymbol{k}_x$) can be written as a product of a Fourier series $\left[\sum_{p'=0}^{\infty} \boldsymbol{c}_{xz,p,p'}(k) \, \Phi_{p'}(z)\right]$ defining the stationary vertical amplitude/phase profile, and a trivial lateral plane wave function $e^{i(kx-\omega t)}$, respectively. Here $\boldsymbol{c}_{p'}$ is the complex vectorial Fourier coefficient of the *p'-th* base function belonging to an arbitrary complete set of $z$-dependent functions $\{\Phi(z)\}$ (see [27] for details). By using the cosine basis functions $\Phi_{p'}(z) = \cos(p' \pi z/d)$, which themself correspond to the profiles of the laterally uniform ($\lambda_{xy} \to \infty$) perpendicular standing spin wave modes (cf. Fig. 1**c**) [34], the individual dynamic magnetization components ($\widetilde{m}_x$ and $\widetilde{m}_z$) of $\widetilde{\boldsymbol{m}}_{xz,p}(x,z,t)$ can be expressed as

$$\widetilde{m}_{x,p,k}(x,z,t) = \left[\sum_{p'=0}^{\infty} \left(a_{p,p'}(k) \cdot \cos(p' \pi z/d)\right)\right] \cdot \cos(kx - \omega t) \qquad (2)$$

$$\widetilde{m}_{z,p,k}(x,z,t) = \left[\sum_{p'=0}^{\infty} \left(b_{p,p'}(k) \cdot \cos(p' \pi z/d)\right)\right] \cdot \sin(kx - \omega t) \qquad (3)$$

where the $k$-dependent coefficients $a_{p,p'}(k)$, $b_{p,p'}(k)$ can be determined as described in Ref. [27]. In the diagonal approximation (i.e. when mode hybridization is neglected) all coefficients with $p' \neq p$ are equal to zero, which makes the vertical profiles essentially independent of $k$ and analogous to those of perpendicular standing waves, yet with finite lateral wave lengths.

So far, it was generally considered that the hybridization of modes of different thickness index was important only in the vicinity of the avoided crossing points of the dispersion curves of these modes, as the hybridization intensity is inversely proportional to the modes' frequency difference. However, analytically calculated mode profiles considering free surface boundary conditions [cf. SM(4b)], that were found to very well match the numerically simulated ones shown here, indicate that a significant hybridization is also present for $k$-values that are well above the avoided crossing points. This strong mode hybridization is a consequence of the fact that the non-diagonal elements of the inter-mode dipole-dipole interaction are strongly $k$-dependent, i.e. they approximately linearly increase with the lateral wave number $k$ [27].

The pronounced hybridization being present at relatively large wave numbers leads to the formation of the rather peculiar mode profile discussed above. Obviously, the flux-closing thickness profile results mainly from the hybridization between the uniform mode ($p=0$) and the *first higher order* mode ($p=1$), with $m_x$ being essentially given by the *first higher order* mode and $m_z$ being dominated by the uniform mode. Although our experiment cannot directly resolve the magnetization distribution along the film thickness, it indirectly confirms the numerically established result that the $m_z$-component has no node, as the presence of a node would lead to a strongly reduced magnetic contrast as opposed to the experimental results.

In summary, we observed the coherent excitation of high-amplitude spin waves with sub-100 nm wavelengths by the driven non-resonant gyration of topologically stabilized magnetic vortex cores in single layer magnetic thin film structures. This excitation mechanism does not require any magnetic bias field, and it was found to persist over a wide frequency range exceeding 10 GHz. We employed time-resolved scanning transmission x-ray microscopy to directly image the corresponding spin-wave propagation, revealing an almost linear dispersion above $k \gtrsim 25$ rad/µm with group velocities of approximately 500 m/s. The experimental results were fully reproduced by micromagnetic simulations, and the observed wave mode was identified as first *higher order* propagating dipole-exchange spin wave mode of the Damon-Eshbach geometry, that was predicted earlier by analytic theory. Remarkably, this *first higher order* mode exhibits substantially lower frequencies than the quasi-uniform Damon-Eshbach mode below a certain wavelength, if the magnetic film is of sufficient thickness [cf. SM (4c)]. Furthermore, we found that because of significant mode hybridization, the thickness profile of the observed mode is heterosymmetric, leading to the formation of flux-closing thus strayfield reducing dynamic magnetic vortices over the film profile as well as to regions with anti-Larmor precession sense. Apart from their fundamental importance, our findings may open a new chapter for magnon spintronics, by meeting the demand for confined, efficient and tuneable sources of coherent ultrashort spin waves in the GHz range.

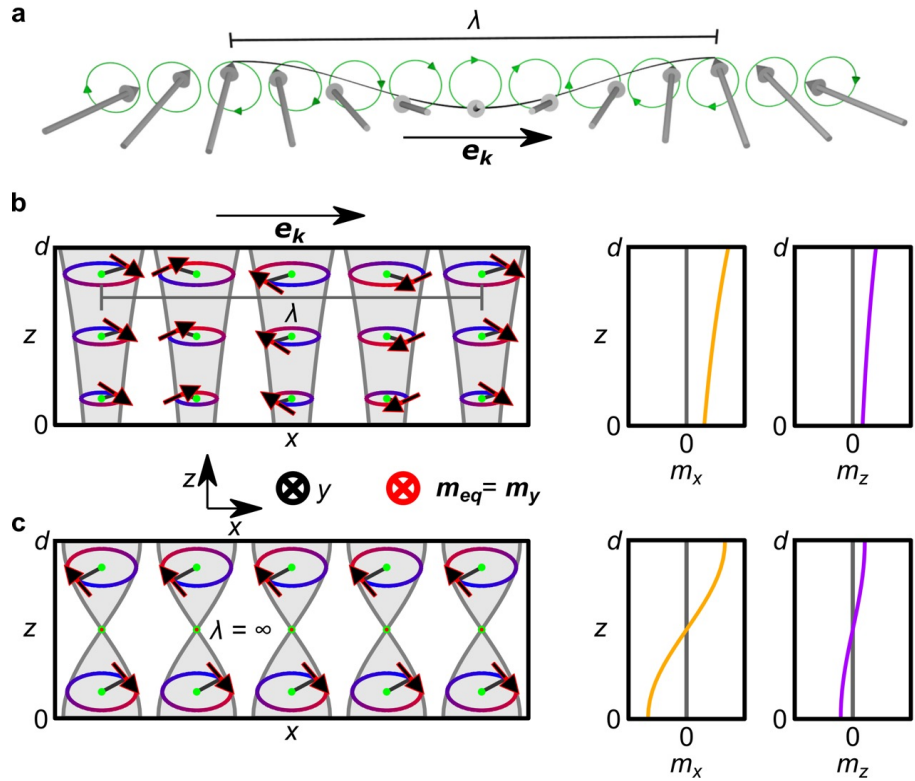

**Figure 1: Schematics of spin waves. a** Spin-wave propagating in the direction of $e_k$. Magnetic moments (grey arrows) precessing with a spatial phase difference determining the wavelength $\lambda$. **b** Cross-sectional precession profile of a quasi-uniform Damon-Eshbach spin-wave in a thin film of thickness $d$, with finite lateral wavelength, propagating along $e_k$. Right panels: Separated profiles of the dynamic $m_x$ (yellow line) and $m_z$ (purple line) components. **c** Cross-sectional precession profile of the first perpendicular standing spin-wave mode for free surface boundary conditions and with laterally uniform magnetization precession (infinite lateral wavelength). Right panels: Separated profiles of the dynamic $m_x$ (yellow line) and $m_z$ (purple line) components.

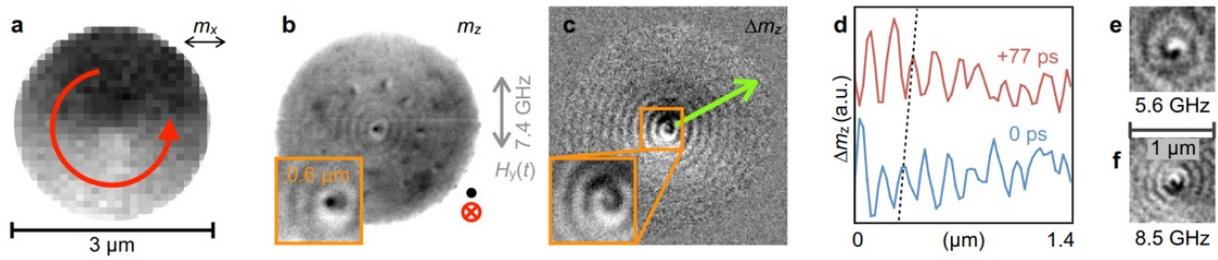

**Figure 2: Emission of ultrashort spin waves from a magnetic vortex core.** (**a**) X-ray microscopy image showing the in-plane magnetic vortex configuration of the disc where the red arrow indicates the orientation of the static magnetization. (**b**) and (**c**) Time-resolved x-ray microscopy images (snapshots) with out-of-plane sensitivity showing the response of the sample to an in-plane alternating magnetic field excitation of $f = 7.4$ GHz. (**b**) Absolute absorption images (full sample and magnified center region) revealing the vortex core in the center of the disc (black dot) and a radial spin-wave pattern with $\lambda = 140$ nm. (**c**) Normalized images showing only the temporal magnetic changes in the disc with respect to the temporally averaged magnetization state, clearly highlighting the spin-wave spiral emitted. (**d**) Line profiles along the green arrow in (**c**) at a relative time delay of 77 ps (from blue curve to red curve) illustrating the high spin-wave amplitudes and their propagation. (**e**) and (**f**) Normalized x-ray images showing $\Delta m_z$ for excitation frequencies of $f = 5.6$ GHz and $f = 8.5$ GHz, respectively.

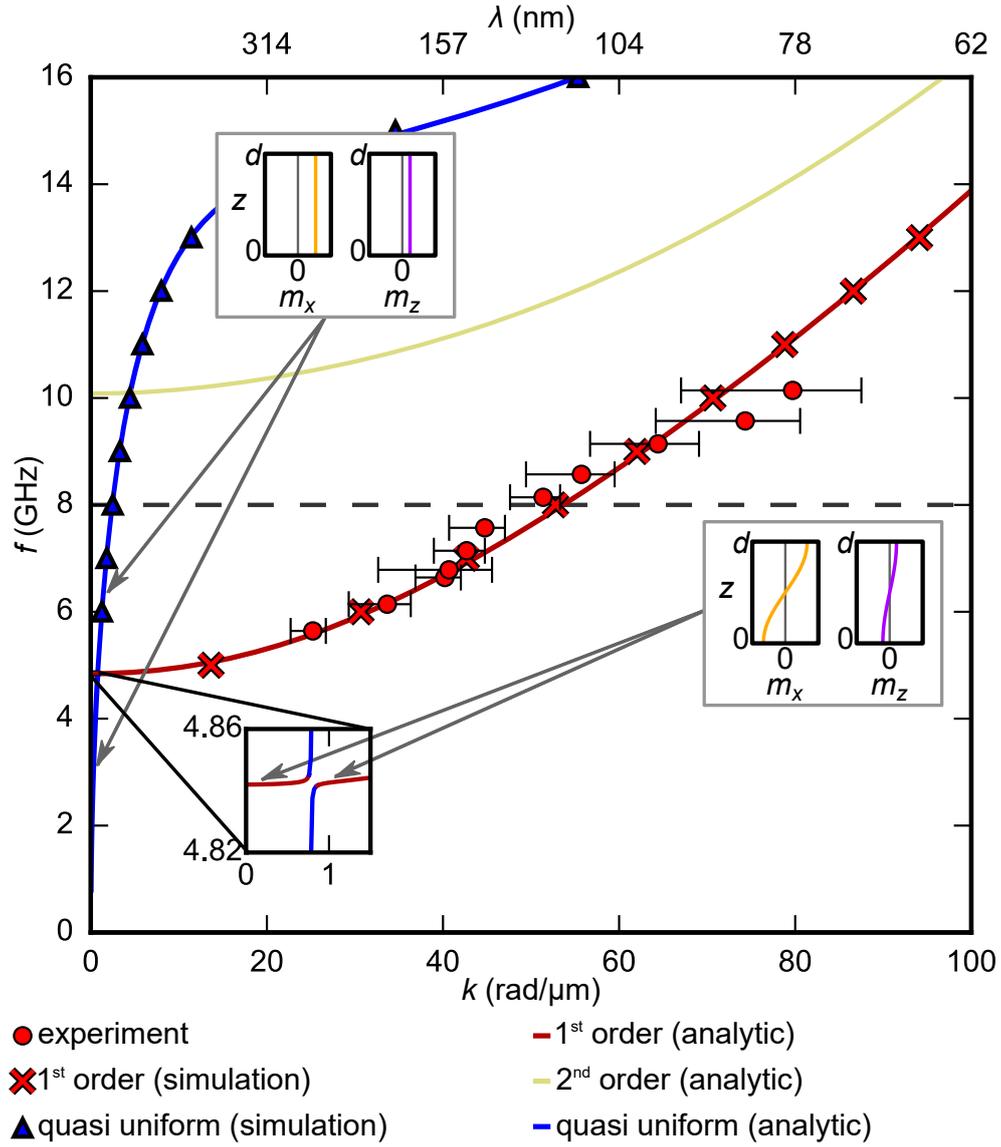

**Figure 3: Spin-wave dispersion relations *f(k)*.** Solid red dots correspond to the experimental data points measured in a single layer vortex structure by time-resolved x-ray microscopy. Solid lines correspond to the analytically calculated dispersion curves of the quasi-uniform branch (blue) and the *first* (red) and *second* (green) *higher order* branches, considering mode hybridization. Results from micromagnetic simulations are marked as blue triangles (quasi-uniform mode) and red crosses (*first higher order* mode). The inset shows the details of the calculated avoided mode crossing of the dispersion curves. Schematic mode profiles for certain dispersion points are shown as yellow (dynamic $m_x$ component) and purple (dynamic $m_z$ component) lines. The dashed horizontal line indicates a frequency of 8 GHz.

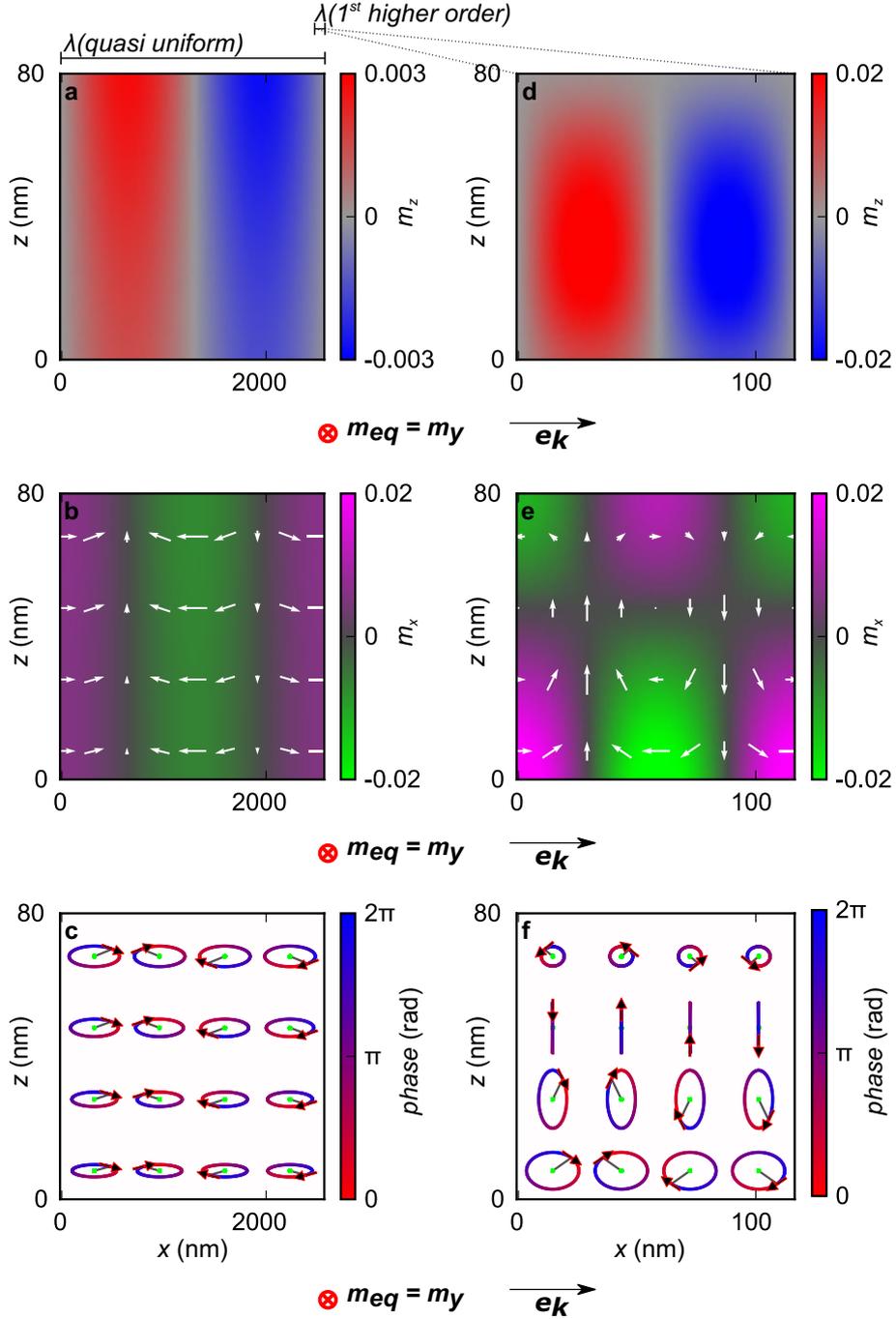

**Figure 4: Spin-wave thickness profiles from micromagnetic simulations at $f$ = 8 GHz.** The left column (**a-c**) shows the quasi-uniform branch, while the right column (**d-f**) shows the first *higher order* branch. The first line (**a,d**) displays a snapshot of the dynamic $m_z$ component as red and blue contrast (spatially smoothed). The second line (**b,e**) shows a snapshot of the dynamic $m_x$ component collinear to $e_k$ as red and green contrast (spatially smoothed), with additional indication of both dynamic magnetic components as white arrows. The third line (**c,f**) shows the position dependent (green points) precession orbit with true aspect ratios as well as the precession sense and phase.

## Acknowledgements

We would like to thank R. Verba and J. Lindner for fruitful discussions, M. Bechtel for experimental support at the MAXYMUS STXM beamline, K. Hahn and U. Eigenthaler for the TEM measurements and R. Mattheis for his help with the sample fabrication. Support by the Nanofabrication Facilities Rossendorf at IBC and by the Laboratory for Micro- and Nanotechnology at PSI is gratefully acknowledged. The experiments were performed at the MAXYMUS endstation at BESSY II in Berlin, Germany. We thank the HZB for the allocation of synchrotron radiation beamtime. S.F. acknowledges funding from the EU Horizon 2020 project MAGicSky (Grant No. 665095). V.T. and A.N.S. acknowledge funding by the grant No. EFMA-1641989 from the National Science Foundation of the USA, by the grant from DARPA, and by the grant from the Center for NanoFerroic Devices (CNFD) and Nanoelectronics Research Initiative (NRI). D.A.B. and H.Yu.M.-S. acknowledge support by the DFG within Spin+X SFB/TRR 173. S.W. acknowledges funding from the European Community's Seventh Framework Programme (FP7/2007-2013) under grant agreement n.°290605 (PSI-FELLOW/COFUND).

# Coherent excitation of heterosymmetric spin waves with ultrashort wavelengths - Supplemental Material

G. Dieterle[1,*], J. Förster[1], H. Stoll[1], A. S. Semisalova[2], S. Finizio[3], A. Gangwar[4],

M. Weigand[1], M. Noske[1], M. Fähnle[1], I. Bykova[1], J. Gräfe[1], D. A. Bozhko[5],

H. Yu. Musiienko-Shmarova[5], V. Tiberkevich[6], A. N. Slavin[6], C. H. Back[4],

J. Raabe[3], G. Schütz[1] and S. Wintz[2,3,*]

---

[1] Max-Planck-Institut für Intelligente Systeme, 70569 Stuttgart, Germany
[2] Helmholtz-Zentrum Dresden-Rossendorf, 01328 Dresden, Germany
[3] Paul Scherrer Institut, 5232 Villigen PSI, Switzerland
[4] Universität Regensburg, 93053 Regensburg, Germany
[5] Technische Universität Kaiserslautern, 67663 Kaiserslautern, Germany
[6] Oakland University, Rochester, MI 48309, USA
* email: dieterle@is.mpg.de; sebastian.wintz@psi.ch

# (1) Samples and methods

## (a) Sample fabrication

The samples were fabricated on x-ray transparent silicon-nitride membranes of 200 nm thickness on a high resistivity silicon frame. $Ni_{81}Fe_{19}$ (Permalloy) thin films were deposited onto such membranes by magnetron sputtering together with an Al capping layer of 5 nm thickness for oxidation protection. The thicknesses of the Permalloy layer was estimated to ~ 80 nm using cross-sectional transmission electron microscopy [cf. 5(b)]. The patterning of discs with 3 μm diameter was achieved by a combination of electron beam lithography and ion beam etching. After an initial oxygen plasma treatment for resist adhesion purposes, a negative resist (MA-N 2910) was spun onto the film. The microelements were then exposed in the resist by electron beam lithography. Subsequently, the samples were developed for 300 s in MA-D 525 and rinsed in de-ionized water. As a final step, the samples were physically etched by an argon ion beam at two different angles (85° and 5°) whereby only the microelements remain from the original film. The remaining resist was removed by rinsing the sample in acetone and by a second oxygen plasma treating. For alternating magnetic field application, a copper strip of 200 nm thickness was fabricated on top of the microelements by means of electron beam lithography, electron beam evaporation deposition, and lift-off processing [S1]. The patterned microstrip has a width of 5 μm, thus the in-plane magnetic Oersted field at the sample position resulting from an electric current $I$ flowing in the strip is approximately $\mu_0 H = 4\pi \cdot 10^{-2}$ mT/mA·$I$.

## (b) Time-resolved scanning transmission x-ray microscopy

The magnetization dynamics in the Permalloy disc was imaged using time-resolved scanning transmission x-ray microscopy [S2][S3] at the MAXYMUS endstation at BESSY II, Berlin. In the MAXYMUS microscope a monochromatic x-ray beam is focused to a spot on the sample using a Fresnel zone plate with an outermost zone width of 18 nm that enables a lateral resolution of about 25 nm. For image acquisition, the sample was laterally translated with respect to the beam and the transmitted x-ray intensity was measured by a fast avalanche photo diode. Magnetic contrast was obtained by tuning the photon energy to the resonant Fe $L_3$ absorption edge (~ 708 eV), at which strong x-ray magnetic circular dichroism [S4] occurs. The x-ray magnetic circular dichroism effect is directly proportional to the magnetization component collinear to the incident photon beam. It therefore allows for a quantitative determination of the local magnetic moment in an element specific manner by probing differences in the transmitted intensity. The out of plane magnetization component was imaged in a normal incidence setup while measurements with oblique incidence were providing additional information on the in-plane magnetization components.

The magnetization dynamics in the Permalloy element was excited by a local magnetic Oersted field. This field was generated via an alternating electric current flowing through a copper stripline on top of the sample. To characterize the electric excitation signal, both the input signal and the transmitted signal were analyzed using power meters and a spectrum analyzer. Taking advantage of the inherent time structure of the BESSY II synchrotron radiation, i.e. x-ray flashes with a repetition rate of 500 MHz, the magnetization dynamics was stroboscopically imaged with a temporal resolution of about 10 ps, determined by the temporal width of the x-ray flashes in the low-alpha operation mode. Given this time structure, the excitation frequency had to be chosen as a rational multiple $(M/N)$ of the 500 MHz probe frequency, where $M$ is a free multiplier and $N$ corresponds to the number of counting registers used. The intensity transmitted for each x-ray flash was detected by the fast avalanche photo diode and the resulting signal was routed to its respective counting register using a field-programmable gate array. Note that for a given excitation frequency $(M/N) * 500$ MHz, also higher integer $(M')$ harmonic responses of the sample can be resolved by this technique as long as $(M' * M) \bmod N \neq M \bmod N$. In case of $M' = 2$ (frequency doubling) this criterion is fulfilled for all experimental data points.

### (c) Micromagnetic simulations

Micromagnetic simulations based on the time integration of the Landau–Lifshitz–Gilbert equation [S5][S6][S7] were carried out using the MuMax3 simulation program [S8]. A cell size of (5 x 5 x 5) nm$^3$ and the following material parameters were chosen for Permalloy: A Gilbert damping constant of $\alpha = 0.007$, a saturation magnetization of $M_s = 800$ kA/m and an exchange constant of $A = 0.75 \cdot 10^{-11}$ J/m$^{-1}$. Intrinsic anisotropies were neglected for symmetry reasons. Note that the value of $A$, resulting from the best fit to the experimental data, is smaller than what is typically assumed for Permalloy thin films, yet it is in agreement with recent reports for as-deposited (non-annealed) films of moderate thickness [S9]. For simulating plane spin waves in a continuous film [S10], a stripe geometry with a length of 40 μm (along $x$), a thickness of 80 nm (along $z$), and a width of 80 nm (along $y$) was modelled. Periodic boundary conditions were set along the $y$ axis which essentially mimics an infinite extension of the film along this axis. The magnetization was initialized to be pointing in the $y$-direction and stabilized by a small external magnetic bias field of $\mu_0 H = 1$ μT. The spin waves were then excited by an alternating perpendicular magnetic field (amplitude $\mu_0 H = 1$ mT) extended over the full (periodically repeated) width of the stripe, yet confined to a narrow *x*-region of 10 nm in the center of the stripe. Simulations were carried out over a wide frequency range. For determining the dispersion relation, $m_z$ was averaged over the sample thickness and then analyzed in the +x half space from the region of the excitation field, by means of Hamming filtering and a subsequent spatial Fourier transform along the x-axis, thus extracting the corresponding wave numbers *k* for a given frequency. For the analysis of the mode profiles, the perpendicular excitation field was modulated with the frequency-correspondent wavelength and it was set to propagate along the *x* direction.

**(d) Analytic calculations**

Analytic calculations were carried out in accordance with the method described in [S11] taking into account the hybridization of the first ten orders of modes to ensure convergence of the results. All calculations were made for the case of free surface boundary conditions (unpinned surface spins).

## (2) Supplemental movies and figures

### (a) Time-resolved scanning transmission x-ray microscopy

Selected time-resolved scanning transmission X-ray microscopy measurements are shown as movies (**M1-M6**) with perpendicular magnetic sensitivity for different excitation frequencies. These movies are labelled according to the type of contrast they show [absolute contrast ($m_z$) vs. normalized contrast ($\Delta m_z$)], their lateral size (µm), the excitation frequency used (GHz) and the corresponding observation period (ps) as follows:

*M#_contrast_size_frequency_period*

**M1**_absolute_3.5µm_7.4GHz_135ps

**M2**_ normalized_3.5µm_7.4GHz_135ps

**M3**_ normalized_0.6µm_7.4GHz_135ps

**M4**_ normalized _0.6µm _7.4GHz_135ps

**M5**_ normalized _1µm_5.6GHz_178ps

**M6**_ normalized _1µm_8.5GHz_118ps

All movies show 10 repetitions of the observation period for a better visibility of the dynamic effects.

### (b) Micromagnetic simulations

The simulated magnetic response of the sample [cf. SI (2)] to an excitation of 7.4 GHz is provided as movie showing the perpendicular magnetic component $m_z$, averaged over the film thickness:

**M7**_absolute_3µm_7.4GHz_135ps_sim

The spin wave mode profile of the first higher order mode, shown in Fig. 4 of the main text, is depicted at subsequent time steps in a movie, where the different panels appear in the order as in Fig. 4 of the main text and a description can be found in the corresponding figure caption:

**M8**_spin_wave_profile_8GHz_sim

For a better visibility of the dynamic effects both movies show 10 repetitions of the excitation period.

**Figure S.1**

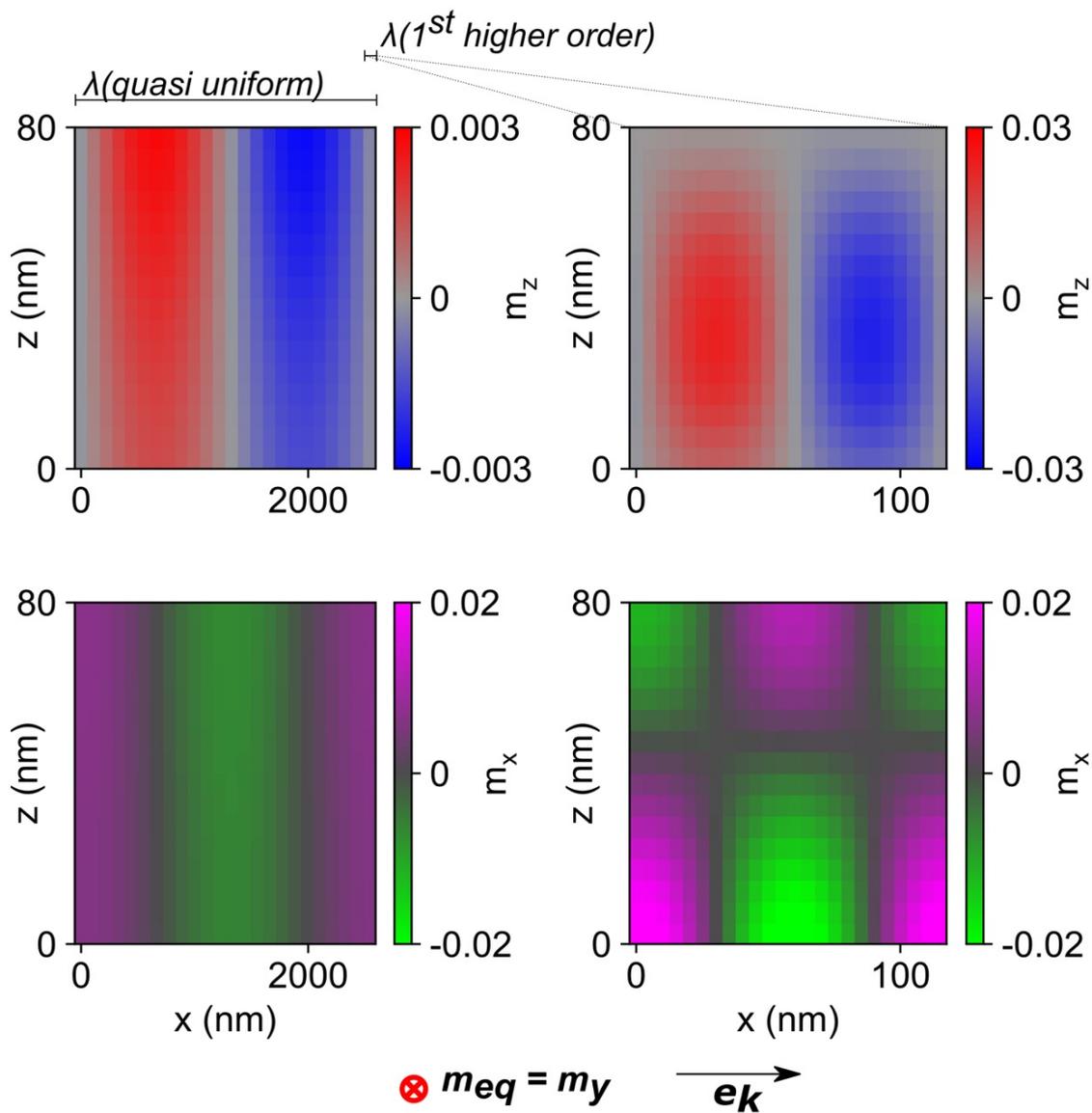

Figure S.1 Non-smoothed spin-wave profiles of the quasi-uniform mode and the first higher mode at an excitation frequency of 8 GHz, revealing the discretization in cells. A description can be found in the caption of Fig. 4 in the main text.

## (3) Supplemental micromagnetic simulations

### (a) Micromagnetic simulations of spiral shaped spin waves in a vortex structure

For a better understanding of the experimental findings, micromagnetic simulations of spin-wave emission in vortex structures were carried out. For this purpose, a disk-shaped grid with a diameter of 3 µm and a thickness of 80 nm (i.e. 600 cells x 600 cells x 16 cells with a cell size of 5 nm x 5 nm x 5 nm) was used. As for the simulations described in the methods section, a Gilbert damping constant of $\alpha = 0.007$, a saturation magnetization of $M_s = 800$ kA/m and an exchange constant of $A = 0.75 \times 10^{-11}$ J/m were chosen and intrinsic anisotropies were neglected. The sample was excited by a homogenous in-plane rotating external magnetic field of 7.4 GHz linear frequency and the field amplitude was gradually increased to 0.1 mT over 10 ns.

Figure S.2 shows the static magnetization distribution of such a vortex structure in cross-sectional view, separately for the three magnetic components $m_x$, $m_y$ and $m_z$. Note that there are comparably small radial magnetisation components arising in the core vicinity with opposing sense in the top and bottom half of the film, as to partially close the flux of the vortex core magnetisation. This magnetisation feature, however, was not observed to influence the spin waves propagating within the disc, in particular with respect to their dispersion relation.

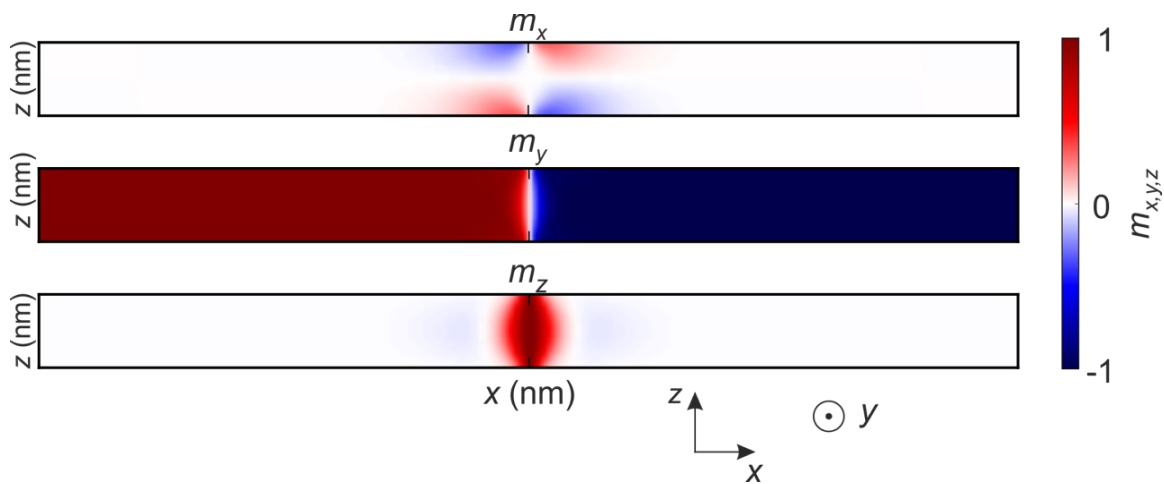

Figure S.2 Simulated static vortex configuration in cross-sectional view.

### (b) Micromagnetic simulations: Continuous film vs. vortex and point source

Figure S.3 shows that the dispersion relation for an infinite film determined by micromagnetic simulations is in very good agreement with both the infinite film analytic theory and the experimental results for a vortex structure. Furthermore, this dispersion relation coincides with simulations of a vortex structure where the magnetization was excited by the vortex core as well as by a point source oscillating out-of-plane field located in the middle between the disc center and the edge. These simulations underline that the thin film approximation is valid for spin waves in vortex structures when the wavelength is much smaller than the structure's diameter.

To obtain the dispersion relation of spin waves in vortex structures, we performed simulations at different excitation frequencies by extracting the corresponding wavelength from different maxima of $m_z$. The dispersion relation for the continuous film was determined as described in the methods section of the main text.

**(c) Size dependence of vortex core driven spin-wave emission**

To investigate the influence of the lateral sample dimensions on the spin-wave emission process, micromagnetic simulations for different disc diameters (1700, 1000, 500, 250 nm) were performed using an in-plane rotating excitation field of $f = 5$ GHz. Here, the exchange constant was set to $A = 0.55 \cdot 10^{-11}$ J/m while a saturation magnetization of $M_s = 827$ kA/m and a thickness of 100 nm was chosen. The results are shown in Figure S.4 as temporal snapshots of the lateral $m_z$ distributions and corresponding line profiles through the centers of the discs. For diameters of the order of 1 μm and above, there is no significant influence of the lateral size: The spiral spin-wave pattern in a smaller disc approximately resembles an equally sized fraction of the spiral in a larger disk. As the spin-wave signal decreases from the center to the edge, reducing the diameter of the disc results in spirals with higher amplitudes reaching the edges. For discs with smaller diameters than 1 μm, the spin-wave pattern is influenced by reflections from the edges. These simulations show, nevertheless, that by using smaller discs, spin-wave amplitudes of $m_z \approx 0.25$ can be achieved at the disc edges. In order to obtain as large spin-wave amplitudes as possible the magnetic field amplitude was chosen to be slightly below the threshold for vortex core switching.

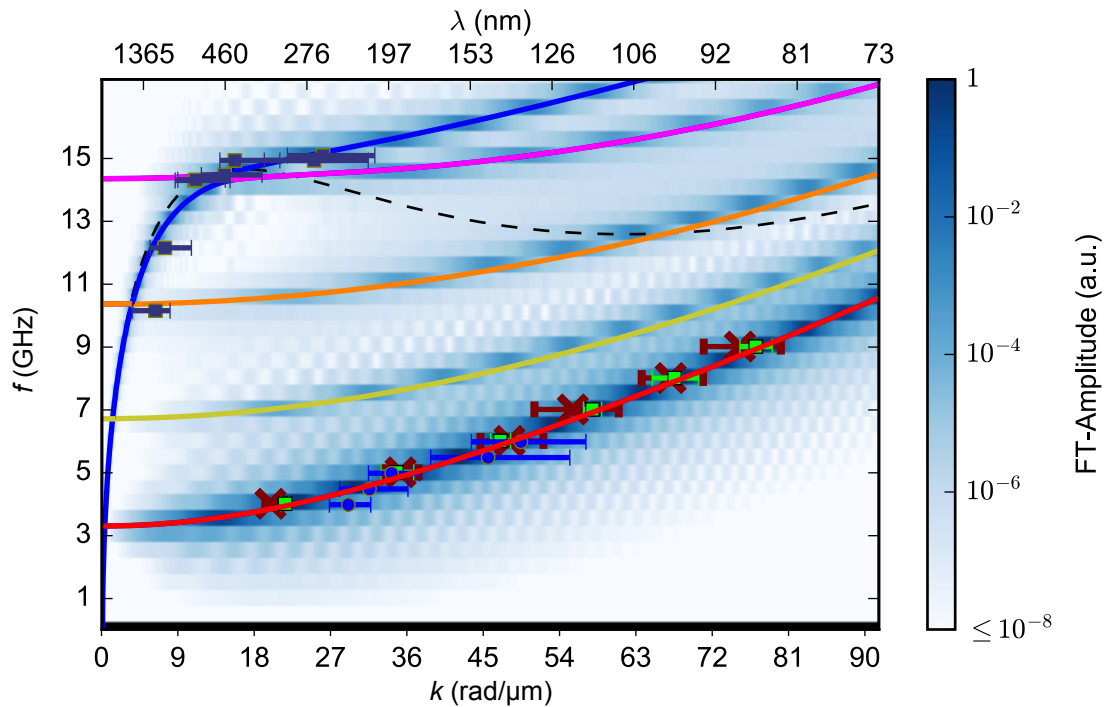

- - - Damon-Eshbach [analytic, diagonal approach (neglecting hybridisation)]
— 1st higher order (analytic)
— 2nd higher order (analytic)
— 3rd higher order (analytic)
— 4th higher order (analytic)
— quasi uniform (analytic)
• 1st higher order (experiment, sample diameter: 1.7µm)
■ 1st higher order excited by homogenous field (simulation, sample diameter: 1.7µm)
✕ 1st higher order excited by point shaped out of plane field (simulation, sample diameter: 1.7µm)
■ Damon-Eshbach (experiment, sample diameter: 5µm)

Figure S.3: Dispersion relations. Continuous thin film dispersion relations determined by micromagnetic simulations (blue-white contrast) and as analytically calculated (solid lines). Simulated dispersion relations in vortex structures, for the sample being excited by a homogenous rotating magnetic field (green squares) or by a locally oscillating perpendicular magnetic field (red crosses). Experimental dispersion relations measured in vortex structures (quasi-uniform mode: deep blue squares, first higher order mode: light blue dots).

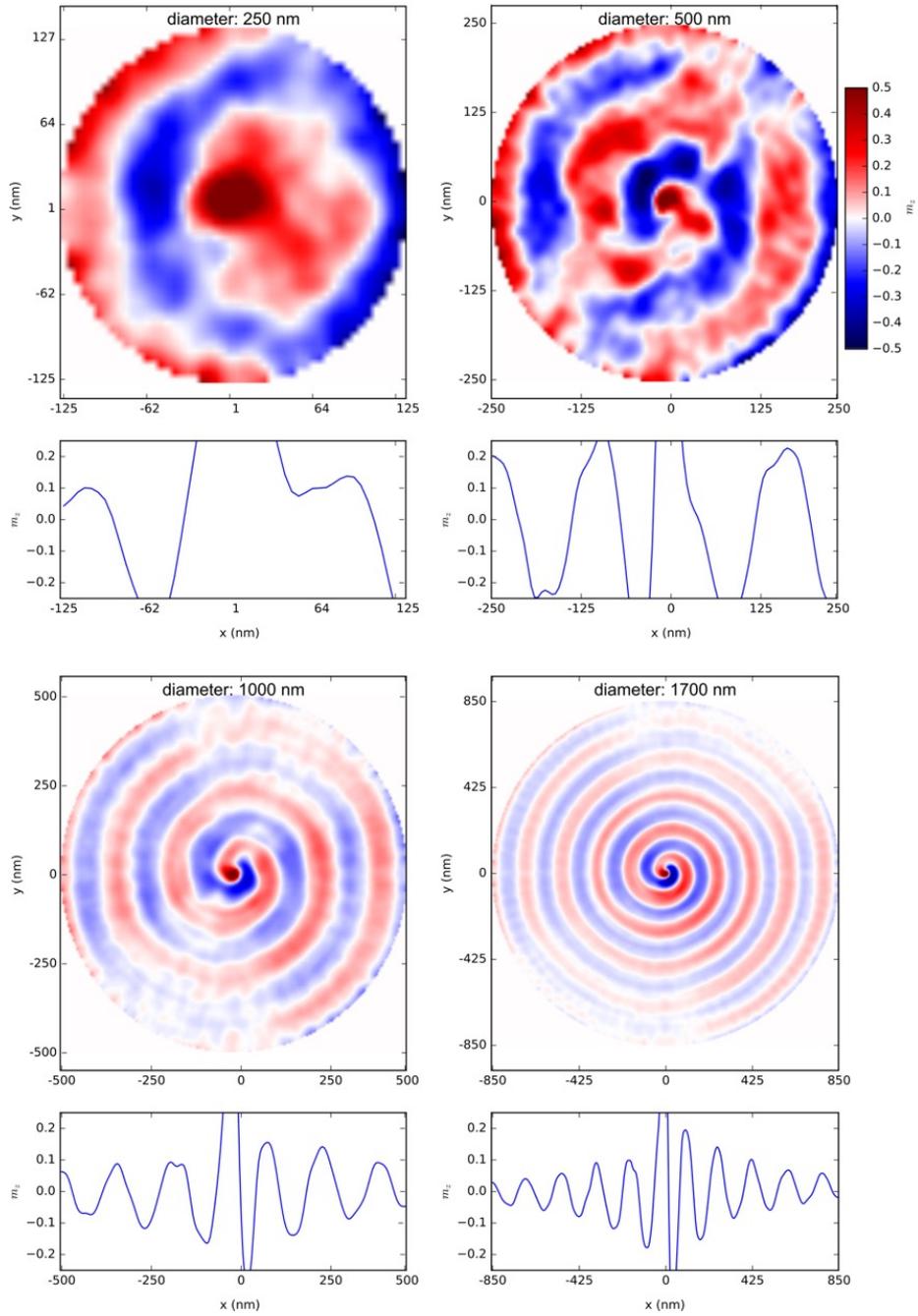

Figure S.4: Representative snapshots of micromagnetic simulations performed for different disc diameters shown as thickness-averaged lateral $m_z$ distribution (red-white-blue contrast) and as corresponding line profiles through the centers of the discs (blue curves). An excitation frequency of 5 GHz and excitation amplitudes (250 nm: 10 mT; 500 nm: 12 mT; 1000 nm: 5 mT, 1700 nm: 3 mT) slightly below the vortex core switching threshold were chosen.

**(d) Vortex core driven spin-wave generation mechanism**

With respect to the origin of the vortex core driven spin-wave generation, it has been a question in which way the spatially almost homogeneous oscillating magnetic field acting on the sample is transformed into spin waves of finite, in particular even ultrashort wavelengths. From a basic perspective, two different excitation mechanisms could be considered here, namely the Schlömann mechanism [S14]-[S18] and Fano resonances [S1]-[S1].

The Schlömann mechanism is based on a spatial variation of the ferromagnetic resonance frequency over the sample, originating from inhomogeneous static or dynamic demagnetizing fields. In this way, a locally increased ferromagnetic resonance may transduce to adjacent regions, where -at the same frequency- the dynamics already correspond to spin waves of finite wavelengths. Analytic approaches to describe this mechanism [S18] require a situation where on a dominant magnetic equilibrium background $\mathbf{M_0}$ small oscillations $\mathbf{\delta M}(t)$ are excited directly by the external magnetic field $\mathbf{H}(t)$.

On the other hand, a Fano resonance exists in general when a system continuous in frequency (here spin-wave dispersion) is coupled to a more discrete resonance (potentially vortex core gyration).

From both, general arguments and micromagnetic simulations we conclude that vortex core driven spin-wave generation falls into the category of Fano resonances rather than of the Schlömann mechanism. This is for the reasons as follow:

From experiments as well as simulations we know that a small but finite core gyration (orbiting of the core around its equilibrium position) coincides with the generation of spin waves. This implies that the magnetic orientation may change up to 180° during each excitation cycle in the core vicinity where the actual excitation occurs. While this point does not rule out a more general Schlömann like mechanism, it considerably complicates such a description. Secondly, we argue that this core gyration is even causal for the spin-wave generation observed as simulations using perpendicular field excitation of the same magnitude as for the in-plane case, -thereby not causing core gyration- did not result in any excited waves at the core, which is in disagreement with the Schlömann mechanism.

In order to identify the details of the spin-wave generation discussed, we performed a corresponding micromagnetic simulation. In the following, for the sake of simplicity, we consider the case of a rotating in-plane field for exciting the circular structure. In the simulation a vortex structure in a disc with a diameter of 1 μm and a thickness of 80 nm (i.e. 200 cells x 200 cells x 15 cells with a cell size of 5 nm x 5 nm x 5.33 nm) was assumed. We only consider the central discretization plane for the discussion, yet the situation in the other planes is qualitatively similar. For the magnetic properties, a Gilbert damping constant of $\alpha = 0.01$, a saturation magnetization of $M_s = 800$ kA/m and an exchange constant of $A = 0.75 \times 10^{-11}$ J/m were chosen. The magnitude of the in-plane rotating field ($f = 8$ GHz) was set to $\mu_0 H_{ext} = 0.5$ mT, where the amplitude was gradually increased over 10 ns and the simulation was continued at the full field magnitude for about 40 ns in order to approach steady state dynamics. At the final state of the simulation, the magnetization, the torque, the effective magnetic field $H_{eff}$ and the externally applied magnetic field were extracted. A vortex core polarization pointing "up" and a counter clockwise rotating field were chosen.

In order to elucidate the mechanism of the spin wave generation, we analyzed the individual sources of the total torque $\tau := -|\gamma|\mu_0(\boldsymbol{m} \times \boldsymbol{H}_{eff})$. According to the Landau Lifschitz Gilbert equation (Ref. [S12]), when damping is neglected, the temporal magnetic dynamics ($d\boldsymbol{m}/dt$) is given by:

$$\frac{d\boldsymbol{m}}{dt} = -|\gamma|\mu_0(\boldsymbol{m} \times \boldsymbol{H}_{eff}) \tag{S.1}$$

and therefore, caused by the torque. Here, $\mu_0$ is the vacuum permeability, $\gamma$ the gyromagnetic ratio and $H_{eff}$ the effective magnetic field.

In our case, the local effective field $\boldsymbol{H}_{eff}$ is composed of three different parts: internal exchange and dipolar fields, as well as the external field. Hence, the resulting total torque can be separated into its components stemming from these three different fields. We calculated separately the torque resulting from the externally applied field as well as the torque originating from the sum of the internal fields (demagnetizing and exchange).

Figure S.5 displays simulation results as snapshots of magnetization and torques in the vicinity of the vortex core during spin-wave generation for a central discretization layer of the structure. In panel **a**, the perpendicular magnetization component $m_z$ is shown as blue-white-red contrast, clearly revealing the position of the counter-clockwise gyrating core. Panel **b** shows the magnitude of the total torque originating from the effective field during core gyration and panel **c** shows the torque related to solely the externally applied rotating field.

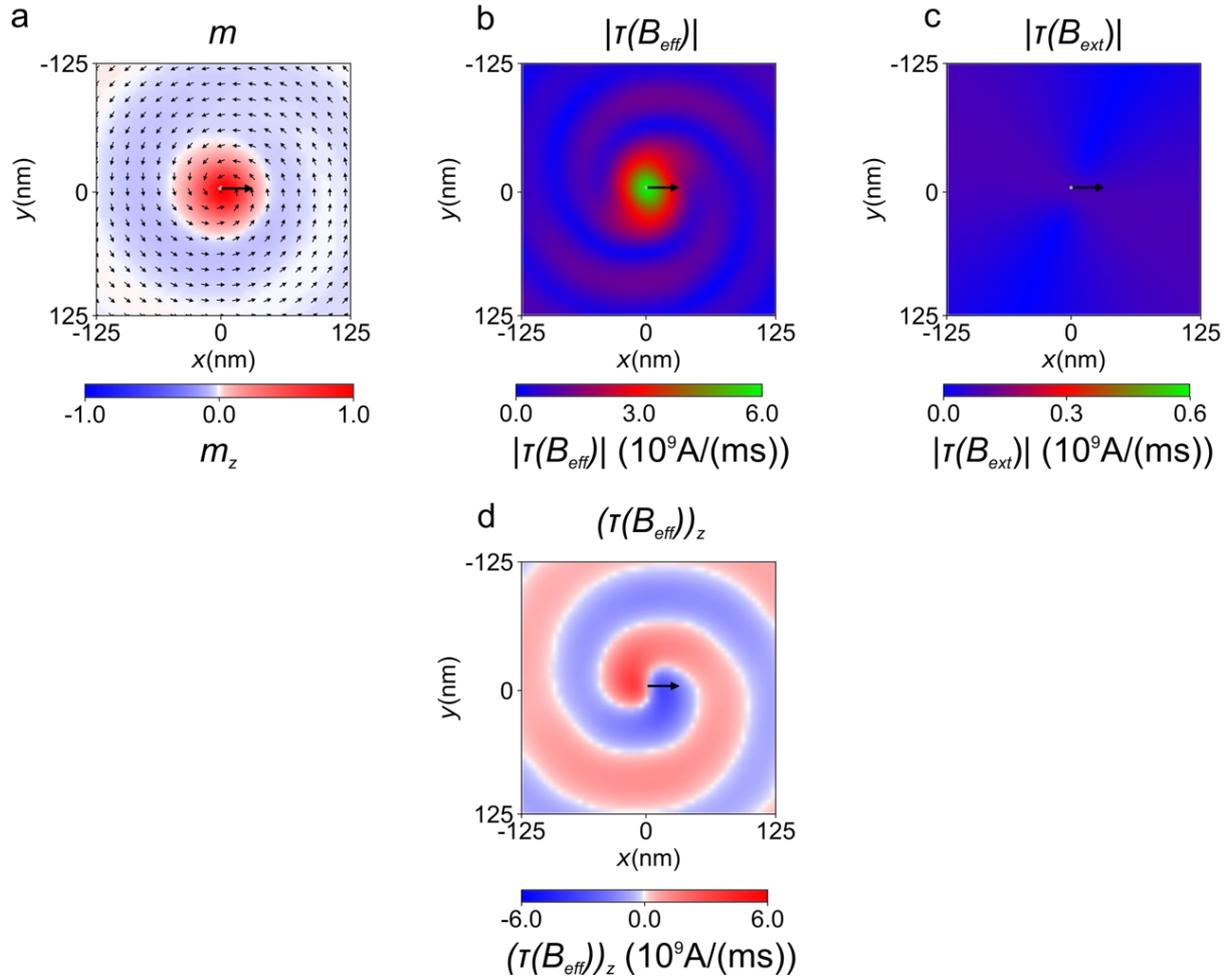

Figure S.5: Simulated snapshots of magnetization and torques during vortex core gyration. (**a**) Perpendicular magnetization component $m_z$ shown as blue-white-red contrast, $m_{xy}$ indicated by small black arrows. (**b**) magnitude of the total torque originating from the effective field (**c**) magnitude of the torque resulting from the externally applied magnetic field; note the color scale difference of a factor of 10. (**d**) $z$-component of the total torque stemming from the effective field. The black arrow indicates the current direction of motion of the vortex core while the grey dot indicates the current core position.

While the external field keeps the gyration to persist, overcoming damping by acting on the whole structure, the local torques in the core vicinity are mainly a consequence from exchange and demagnetizing fields occurring during core gyration. This becomes obvious by comparing the absolute values of the total torque (**b**) to the torque stemming from the external field directly (**c**) where the latter is, in the center region, more than an order of magnitude smaller. As the energy coupled into the spin-wave mode is continuously flowing away from the center as a consequence of wave propagation, a resonant direct field excitation can be ruled out to lead to such a strongly enhanced total torque. In particular, as the magnitude of this torque reaches almost its full value already during the first cycle of excitation, when the field is simulated to start with $\mu_0 H_{ext} = 0.5$ mT from the beginning.

Panel (**d**) shows the *z*-component of the torque originating from the effective field during core gyration. There is, clearly visible, a bipolar structure with opposing sign at diametric sides of the vortex core. This bipolar structure is identified as the excitation source of the spin waves and it is originating from the motion of the vortex core. Hence, an indirect spin-wave excitation mechanism seems natural.

To reveal the effect of the dynamics of the gyrating vortex core, we calculated the gyrotropic field that neglecting damping is given by [S13]:

$$\boldsymbol{H_{gyro}} = -\frac{1}{|\gamma|\mu_0}\left(\boldsymbol{m} \times \frac{d\boldsymbol{m}}{dt}\right) \quad (S.2)$$

If the amplitude is not chosen too high, i.e. below the vortex core switching threshold, the magnetisation is not changing with time in a reference frame that is rotating together with the external field around the axis of the disc. In this rotating reference frame, also $\boldsymbol{H_{eff}}$ and $d\boldsymbol{m}/dt$ are constant in time ($d\boldsymbol{m}/dt$ has to be evaluated incorporating the transformation and is related to the deviation of the magnetisation from the rotational symmetry and the excitation frequency *f*).

Defning an effective field $\widetilde{\boldsymbol{H_{eff}}}$, containing the gyrofield, by:

$$\widetilde{\boldsymbol{H_{eff}}} \equiv \boldsymbol{H_{eff}} + \boldsymbol{H_{gyro}} \quad (S.3)$$

the LLG equation (S.1) can, neglecting damping, be rewritten in the following form:

$$0 = -|\gamma|\mu_0\left(\boldsymbol{m} \times \widetilde{\boldsymbol{H_{eff}}}\right) \quad (S.4)$$

Eq. (**S.4**) has the form of the LLG in the static case, the influence of the dynamics being included in the effective field $\widetilde{\boldsymbol{H_{eff}}}$. Thus in the steady state motion, $\boldsymbol{m}$ would be aligned with $\widetilde{\boldsymbol{H_{eff}}}$ (which in turn depends on $\boldsymbol{m}$ and *f*). Therefore within this rotating reference frame, not the torques but the fields are relevant.

The gyrotropic field calculated from the micromagnetic simulation is shown in Figure S.6 (**b**) and (**c**) for a central discretisation layer of the film. To further illustrate the excitation mechanism we also calculated the gyrotropic field [cf. equation (S.2)] of a gyrating simplified two-dimensional vortex like configuration [cf. Figure S.6 (**d**)]. For this calculation a simplified vortex structure was displaced from the center, the displacement being equal to the radius of gyration in the micromagnetic simulation, and $d\bm{m}/dt$ was calculated assuming this displaced structure to rotate around the centre with a frequency of 8 GHz. The influence of the gyrotropic field on the magnetisation was not taken into account. The perpendicular and in-plane components of the gyrotropic field of this simplified structure are shown in Figure S.6 (**e**) and (**f**), respectively.

The gyrotropic fields calculated from the simplified model and from the micromagnetic simulation are in good agreement in the vicinity of the vortex core. In the gyrotropic field calculated from the micromagnetic simulation, a pair of regions with opposite out of plane components is present at diametric sides of the core. In particular, this bipolar structure is qualitatively and quantitatively reproduced in the simplified model, clearly revealing that the structure originates from the motion of the vortex core.

This bipolar gyrotropic field structure is the source of the bipolar torque structure discussed above (cf. Figure S.5 **d**) as the deformation of the magnetisation from the rotational symmetry directly results in a temporal change of the magnetization. Remarkably, the gyrotropic field is enhanced by a factor of more than 80 compared to the externally applied field in the vicinity of the core. This magnitude of the gyrotropic field along with its strong spatial confinement allows for a high amplitude excitation of spin waves with ultrashort wavelengths. Note that in the simplified model (**d-e**) no spirally shaped structures are present as those are the result of the mutual interaction of the gyrotropic field and the magnetization, which was not taken into account here. Moreover it is important to note that in the simplified model no externally applied field is present.

From the arguments provided above we conclude that the high-amplitude spin-waves, observed in our experiment, are not directly excited by the external field in a Schlömann sense, but are rather a consequence of the dynamics of the field driven vortex core gyration. More abstractly, the effect can be seen as a local Fano resonance of the discrete core gyration mode with the continuum of propagating spin waves, mediated via a coherent and linear coupling.

In contrast to that, the quasi-uniform spin waves excited at the rim of the disc in our experiment cf. S4(a), are solely a consequence of the Schlömann mechanism.

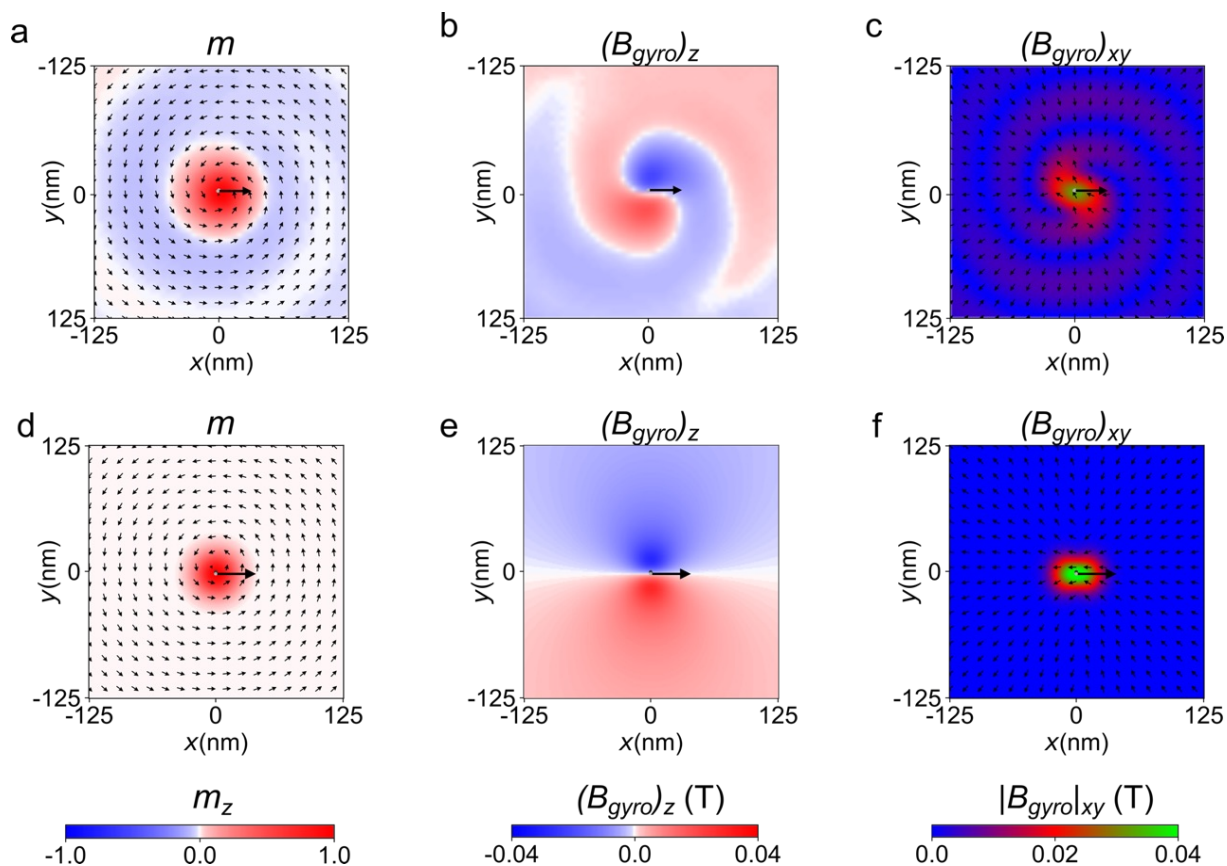

Figure S.6 Simulated snapshots of magnetization and gyrotropic fields. **(a)** Magnetic orientation of a gyrating vortex core with spin waves in the micromagnetic simulation **(b)** z-component of the corresponding gyrotropic field $(B_{gyro})_z$. **(c)** in-plane components of the corresponding gyrotropic field $|B_{gyro}|_{xy}$, where arrows indicate the direction and the magnitude is given by the color scale. **(d)** Magnetic orientation of an approximated two-dimensional vortex configuration with **(e)** $(B_{gyro})_z$ and **(f)** $|B_{gyro}|_{xy}$. The large black arrow indicates the direction of motion of the vortex core and the grey dot its current position in all panels.

**(e) Linearity of vortex core driven spin-wave emission**

To investigate the dependence of the excited spin-wave amplitude on the magnitude of the external driving field, we determined the response of an exemplary disc with 1000 nm diameter (i.e. 200 cells x 200 cells x 16 cells with a cell size of 5 nm x 5 nm x 5 nm) defined in 2(a) as a function of the magnitude of a circularly rotating excitation field (core polarization along +$z$, clockwise field rotation). Except for the excitation field, the simulation parameters are equal to those mentioned in section 2(d). We found the excited spin waves to be coherent up to the field that causes dynamic vortex core switching (switching threshold). Below this field, the excited spin-wave amplitude scales almost linearly with the excitation field up to $\mu_0 H \sim 1.5$ mT, as representatively shown in Figure S.7, where the resulting spin-wave amplitude is plotted for different excitation fields at a position 350 nm off the disc center. In order to obtain this amplitude, a Fourier-transform of the temporal evolution of $m_z$ at a given point was performed for all 16 cell layers. The amplitudes of the Fourier component corresponding to the excitation frequency and its first harmonic were extracted for each layer and then averaged over the dot thickness. We did not observe any noticeable onset threshold field for the process of vortex core driven spin-wave emission. Besides the fundamental frequency, there is also a relatively small spin-wave response of doubled frequency, which is negligible, however, below a field amplitude of $\mu_0 H \sim 0.5$ mT.

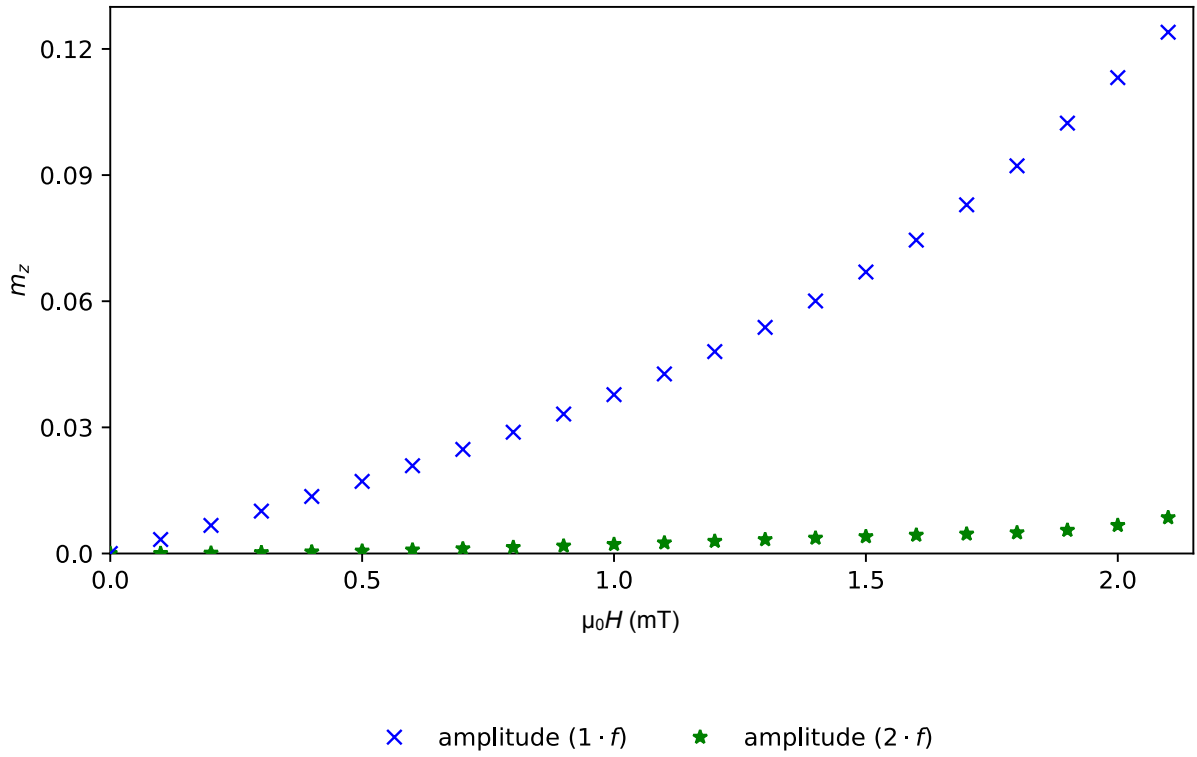

Figure S.7: Simulation of the resulting spin-wave amplitude in dependence of the excitation field for vortex core driven spin-wave emission. The diagram is showing the fundamental (blue crosses) and double frequency (green asterisks) spin-wave amplitude ($m_z$) data points as a function of the excitation field ($\mu_0 H$) at a position 350 nm off the disc center.

## (4) Supplemental analytic spin wave dispersion formulae

### (a) Dispersion formulae

The spin-wave dispersion relations $f(k)$ for an infinitely extended, homogeneously magnetized thin film can be analytically described according to Ref [S11]. Thereby the approximate non-hybridized dispersion of the quasi Damon-Eshbach mode is given by:

$$f(k) = \frac{\gamma \mu_0}{2\pi} \sqrt{\left(H + \frac{2A}{\mu_0 M_S} k^2 - M_S\right) \cdot \left(H + \frac{2A}{\mu_0 M_S} k^2 - M_S \frac{1 - e^{-k \cdot d}}{k \cdot d}\right)}. \quad (S.5)$$

For the case of unpinned surface spins, the dispersion relations of the higher order modes are, neglecting mode hybridization, approximately given by the simplified equation [S11]:

$$f(k) = \frac{\gamma \mu_0}{2\pi} \sqrt{\left(\frac{2A}{\mu_0 M_S}\left(k^2 + \left(\frac{p\pi}{d}\right)^2\right)\right)\left(\frac{2A}{\mu_0 M_S}\left(k^2 + \left(\frac{p\pi}{d}\right)^2\right) + M_S\right)} \quad (S.6)$$

with $p>0$ being a positive integer corresponding to the order of the perpendicular standing spin wave mode. From this equation, the influence of the layer thickness as well as the exchange constant and the saturation magnetization can be estimated.

For $\frac{2A}{\mu_0 (M_S)^2}\left(k^2 + \left(\frac{2\pi}{2d}\right)^2\right) \ll 1$, the first order mode can be further approximated to

$$f(k) \approx \frac{\gamma \mu_0}{2\pi} \sqrt{\left(\frac{2A}{\mu_0}\left(k^2 + \left(\frac{\pi}{d}\right)^2\right)\right)}, \quad (S.7)$$

which is valid for our experiments as $\frac{2A}{\mu_0 (M_S)^2}\left(k^2 + \left(\frac{\pi}{d}\right)^2\right) \lesssim 0.03$.

Obviously, in this approximation, the dispersion does not depend on $M_s$ and the frequency is proportional to $\sqrt{A}$. At the same time, the dispersion strongly depends on the layer thickness for $k \lesssim 1/d$.

All higher order modes exhibit a frequency gap, i.e. no spin waves exist in their corresponding dispersion branch below the minimum frequency $f(0)$ of infinite lateral wavelength ($k = 0$) that can be approximated to

$$f(0) \approx \frac{\gamma p}{2d} \sqrt{2A\mu_0}, \quad (S.8)$$

using a similar inequality $\left(\frac{2A}{\mu_0 (M_S)^2}\left(\frac{2\pi p}{2d}\right)^2 \ll 1\right)$ as for equation (S.7).

The analytic dispersion relations shown in Fig. 3 (main text), Figure S.9 and Figure S.3 respectively were calculated according to the more elaborate method described in Ref [S11] employing a hybridizing set of ten modes. For highlighting the potential inaccuracies of dispersion relations calculated neglecting such hybridization for higher *k*-values, we also show the dispersion curve of the quasi-uniform (Damon-Eshbach) mode, calculated using eq. (S.5) as dashed black line in Figure S.3. Obviously, this non-hybridized dispersion curve significantly deviates from the hybridized one (blue solid line) for $k \gtrsim 20$ rad/μm.

**(b) Analytically calculated mode profiles**

Mode profiles analytically calculated for selected *k*-values of the different hybridized dispersion branches of Fig. 3 (main text) are shown in Figure S.8 together with the corresponding dispersion relations. In general, there is a very good agreement of such calculated profiles with those extracted from the micromagnetic simulations. It should be noted though that the analytical model is restricted to discrete changes of the mode character over the film thickness only (inversion of the binary phase or of the precession sense), making it different to the micromagnetic simulations for which in principle also a continuous change of phase over the thickness could be resulting, yet was not observed in our work.

Near the avoided crossing of the first order mode and the quasi-uniform mode, the mode profiles are practically symmetric, exhibiting one node in both the dynamic $m_x$ and $m_z$-component for the first higher order mode. Towards higher *k*-values, with increasing distance from the avoided crossing point, the nodes in both components gradually shift upwards, yet the node in the $m_z$-component is shifting to a higher extend and it is eventually expelled from the film. This leads to the heterosymmetric mode profile discussed. For even higher *k*-values along the first higher order dispersion branch, also the node in the $m_x$-component is expelled from the film, leading to again to a node-free, somewhat uniform mode profile. The mode profile of the second order mode exhibits a similar behavior, albeit, in the considered range of *k*, with an additional node in each component. Aside from the avoided crossing points with the first two higher order modes, the profile of the zeroth mode is found to be quasi-uniform in phase up to *k* values of ~25 rad/μm. Nevertheless, a significant inhomogeneity in amplitude is exhibited for $k > 20$ rad/μm. This increasing inhomogeneity (or bulging) eventually leads to the simultaneous occurrence of two precessional nodes in each dynamic component within the *k*-range of 25 rad/μm < *k* < 50 rad/μm. This double node formation, however, arises at different *k*-values for the lateral dynamic $m_x$-component (lower *k*) than for the perpendicular dynamic $m_z$-component (higher *k*).

### (c) Thickness dependence of the dispersion relation

In order to evaluate the dependence of the spin-wave dispersion relations on the film thickness, we analytically calculated such dispersions and corresponding mode profiles at selected $k$-values (0.01, 2, 5, 10, 20, 40, 60, 80, 100) rad/µm exemplary for Permalloy layer thicknesses of (25, 50, 100 and 200) nm. From the results of these calculations the following qualitative conclusions can be drawn:

Pronounced heterosymmetric mode profiles only occur when the first higher order mode falls below the quasi-uniform mode in frequency at finite $k$-values. According to our calculations, this transition occurs between a layer thickness of 25 and 50 nm (cf. Figure S.9).

Analytically, the critical transition thickness ($d_{crit}$) can be estimated by equating the asymptotic Damon-Eshbach frequency ($f_{k \to \infty} = \gamma \mu_0 M_s / 4\pi$) (neglecting exchange) with the approximated ferromagnetic resonance frequency [$f(k=0)$] of the first higher order mode (S.8), leading to the following expression:

$$d_{crit} \approx 2\pi \sqrt{\frac{2A}{\mu_0 M_s^2}} = 2\pi l_{ex} \tag{S.9}$$

with $l_{ex}$ being the exchange length (here 4.32 nm), leading to a $d_{crit}$ of ~ 27 nm for Permalloy with the given exchange constant of $A = 0.75 \times 10^{-11}$ J/m.

While heterosymmetric mode profiles with unequal precessional node numbers in the two dynamic magnetization components do not occur below $d_{crit}$, there can still be effects from mode hybridization for smaller thicknesses. For example, a slight shift of the node positions away from the film center (non-uniform for the two dynamic components) can be observed at finite $k$-values for the first higher order mode profile in the 25 nm thick film, see Figure S.9.

Above the critical thickness, heterosymmetry of the first higher order mode profile (no node in $m_x$, one node in $m_z$) occurs at smaller $k$-values when the film thickness is increased. In this way, heterosymmetry arises between (20 and 40) rad/µm for $d = 50$ nm, between (10 and 20) rad/µm for $d = 100$ nm, and already between (2 and 5) rad/µm for $d = 200$ nm. The reason for this trend is that, for higher thicknesses, the quasi-uniform mode 'crosses' the first higher order mode already at smaller $k$-values, which is mainly because of the approximately inverse thickness dependence of the higher order mode frequencies, but also because of a steeper slope at the onset of the dispersion relation of the quasi-uniform mode towards higher thicknesses. Besides this, the dispersion relations for thicknesses above $d_{crit}$ show an increasing number of modes of higher order ($p \geq 1$) falling below the quasi-uniform mode for an increasing film thickness ($p = 1$ for 50 nm, $p = 1, 2, 3$ for 100 nm, and $p = 1...$ 6 for 200 nm).

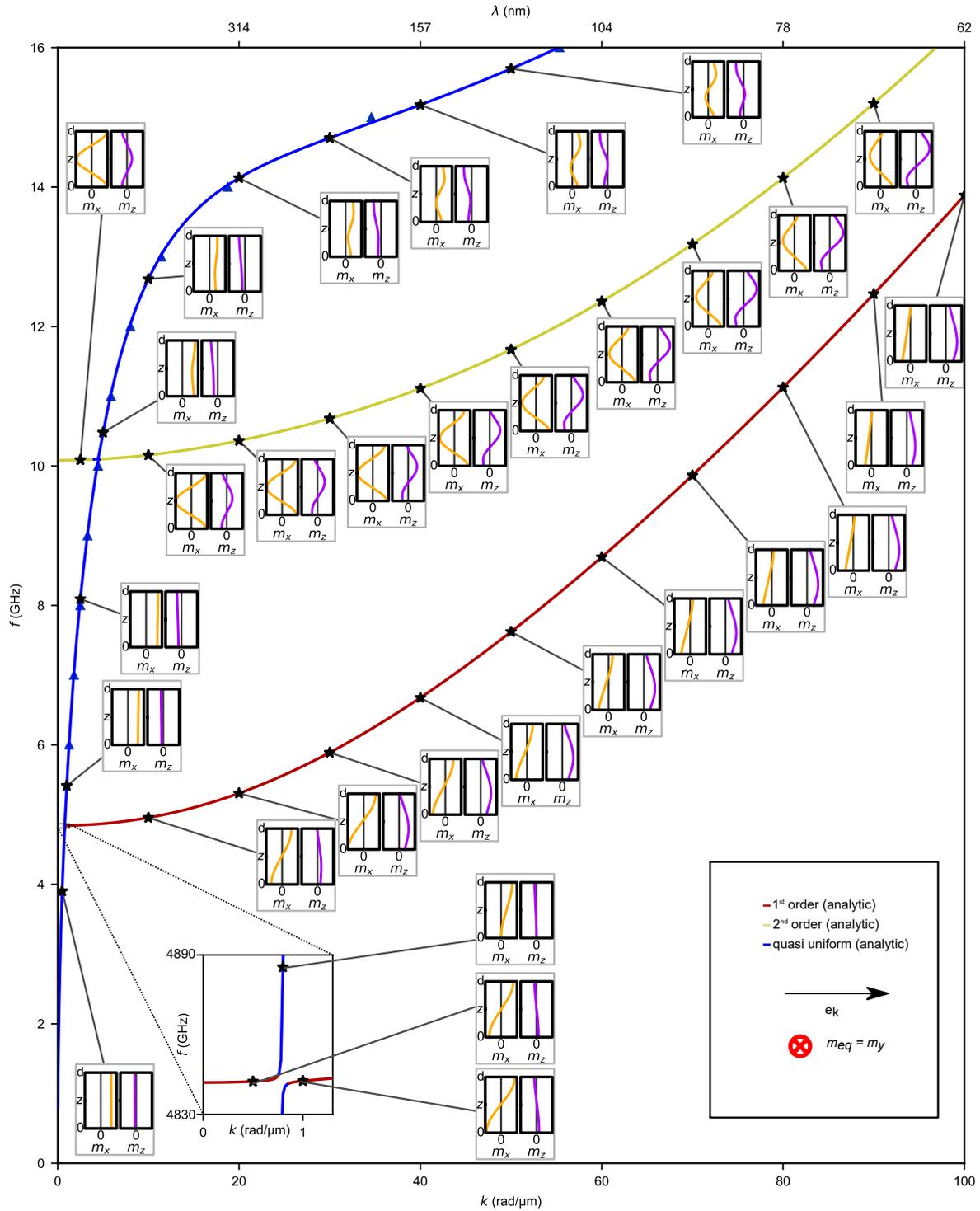

Figure S.8: Analytically calculated mode profiles according to the conditions mentioned in the main text, namely layer thickness $d$ = 80 nm, $M_s$ = 800 kA/m and an exchange constant of $A = 0.75 \times 10^{-11}$ J/m. The solid lines correspond to the analytic dispersion relations as indicated by the legend. The profiles are depicted for the frequency and wavenumber marked by the respective black star. The inset shows the avoided crossing of the quasi uniform mode and the first order mode.

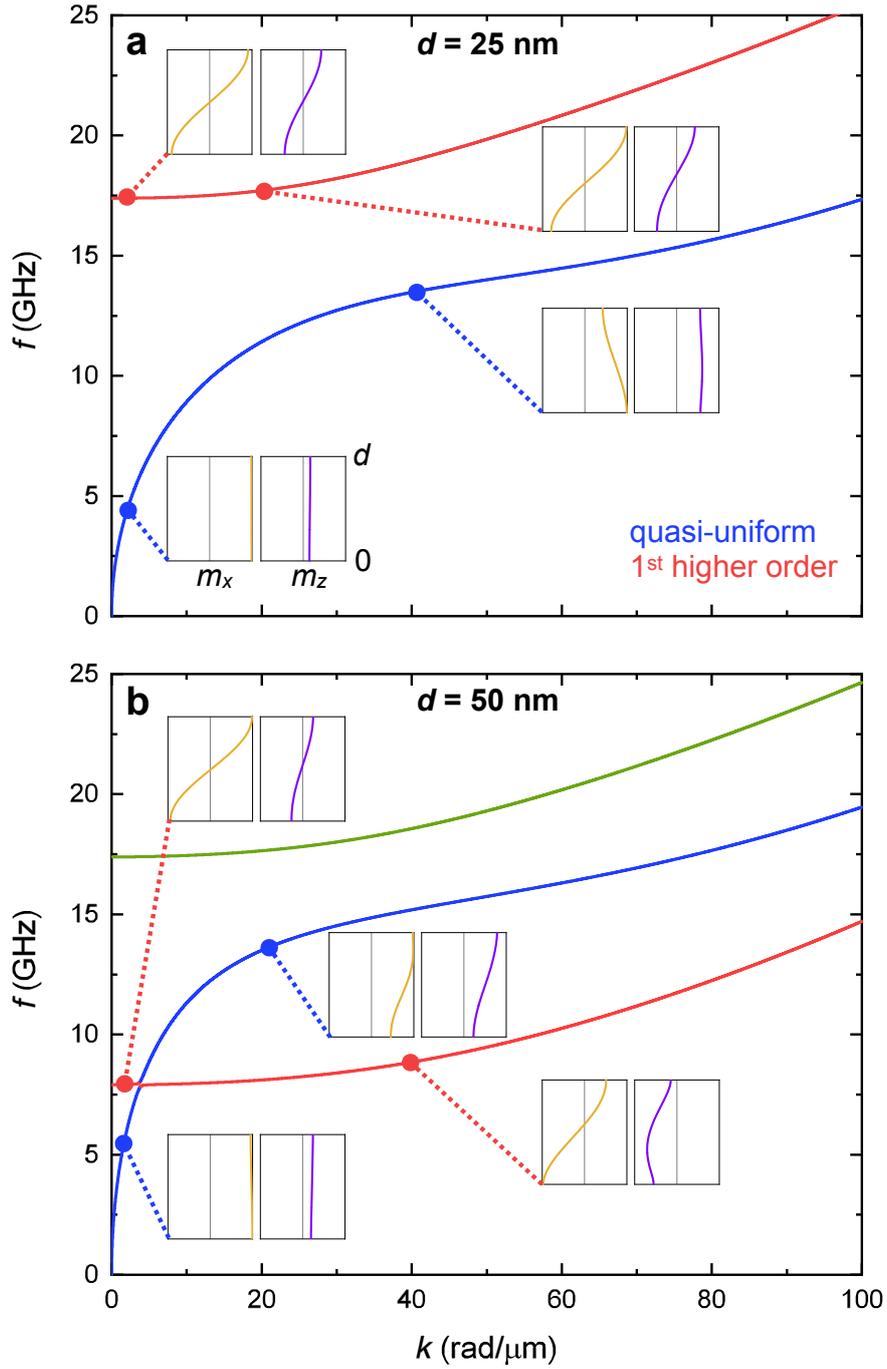

Figure S.9: Spin-wave dispersion relations of the quasi-uniform (blue) and the first higher order mode (red) as analytically calculated for **(a)** 25 nm and **(b)** 50 nm thick Permalloy layers. Cross-sectional dynamic mode profiles are shown for selected $k$-values in yellow ($m_x$) and purple ($m_z$).

## (5) Supplemental Experiments

### (a) Quasi-uniform and first higher order mode

Additional TR-STXM measurements were carried out in order to image both the quasi-uniform mode and the first higher order mode for a common layer thickness. The first higher order propagating spin-wave mode was measured in a 100 nm thick Permalloy sample with a diameter of 1.7 μm. Measurements of the quasi-uniform mode were performed on a larger sample with 5 μm in diameter of a different 100 nm Permalloy deposition tool. While the first higher order spin waves were generated in the center utilizing the driven gyration of the vortex core, quasi-uniform waves were excited using the magnetic discontinuity at the edges of the disc as local perturbation source [S16]. The results are displayed in Figure S.3 as data points of the dispersion relation $f(k)$.

Note that in general both spin-wave modes can be excited in each of the two different discs. This becomes apparent e.g. by means of the movie M2. In addition to the short-wavelength first higher order waves propagating outwards, also a signature of an inwards propagating spin-wave belonging to the quasi-uniform mode can be seen here, yet with a wavelength exceeding the diameter of the disc. Therefore, a larger disk was imaged for measuring the dispersion of the long-wavelength quasi-uniform mode. The excitation of both spin-wave modes did not noticeably depend on the adjacent under- or overlayers of different discs.

Together with the experimental dispersion relation of the quasi-uniform (deep blue squares) and first higher order mode (light blue dots). Figure S.3 also shows the results from micromagnetic simulations (white-blue contrast) as well as analytic calculations (colored solid lines) for a continuous Permalloy film of 100 nm thickness including all spin-wave modes in the Damon-Eshbach geometry up to 17 GHz. Obviously, the results from all three origins are in excellent agreement with each other. Compared to the Permalloy layer with 80 nm thickness, the dispersion curves of two additional higher order modes [$p = 3$ (orange line) and $p = 4$ (pink line)] are situated, for a certain $k$-range, below that of the quasi-uniform mode.

It is noteworthy that first higher order waves are not efficiently excited from the magnetic discontinuity at the edge of the disc. While waves of very low amplitude can be identified in micromagnetic simulations, their amplitude is clearly insufficient to be detected in the experiment. The reason for this low excitation efficiency presumably lies in the mode versus excitation symmetry. The vertical profile of the in-plane component of the first higher order waves is almost antisymmetric whereas the field excitation is homogeneous in space, which from a symmetry point of view points to a strong suppression of the excitation.

On the other hand, in principle, quasi-uniform waves could be excited along with higher order waves from the high-amplitude perpendicular component of the gyrotropic field provided by the gyrating structure in the center. However, at the frequencies, where first higher order waves are emitted, from the vortex core region, the wavelengths of quasi-uniform waves ($\lambda > 1$ μm) are exceeding the antenna size (~core diameter) by almost two orders of magnitude, which in turn renders the excitation efficiency to be considerably reduced. Moreover, when considering quasi-uniform wavelengths approaching the diameter of the core antenna for frequencies above 10 GHz, micromagnetic simulations reveal that the gyration response of the core at such frequencies is too small to cause an experimentally observable spin wave excitation.

**(b) Permalloy layer thickness**

In order to determine the layer thickness of the Permalloy sample discussed in the main text, a bright field transmission electron micrograph of a cross-sectional lamella, extracted from a reference sample was acquired (cf. Figure S.10). Thereby the Permalloy layer thickness was measured to be approximately $d = 80$ nm.

This result supports our assumption of a relatively small exchange constant ($A = 0.75 \cdot 10^{-11}$ J/m) in the Permalloy sample investigated. According to eq. (S.7), for the given experimental conditions, only $d$ and $A$ remain as relevant parameters for the first higher order spin-wave dispersion relation. Since $d$ was directly measured to 80 nm, $A$ has to take the value assumed for the analytic dispersion relation to match with the experimental results.

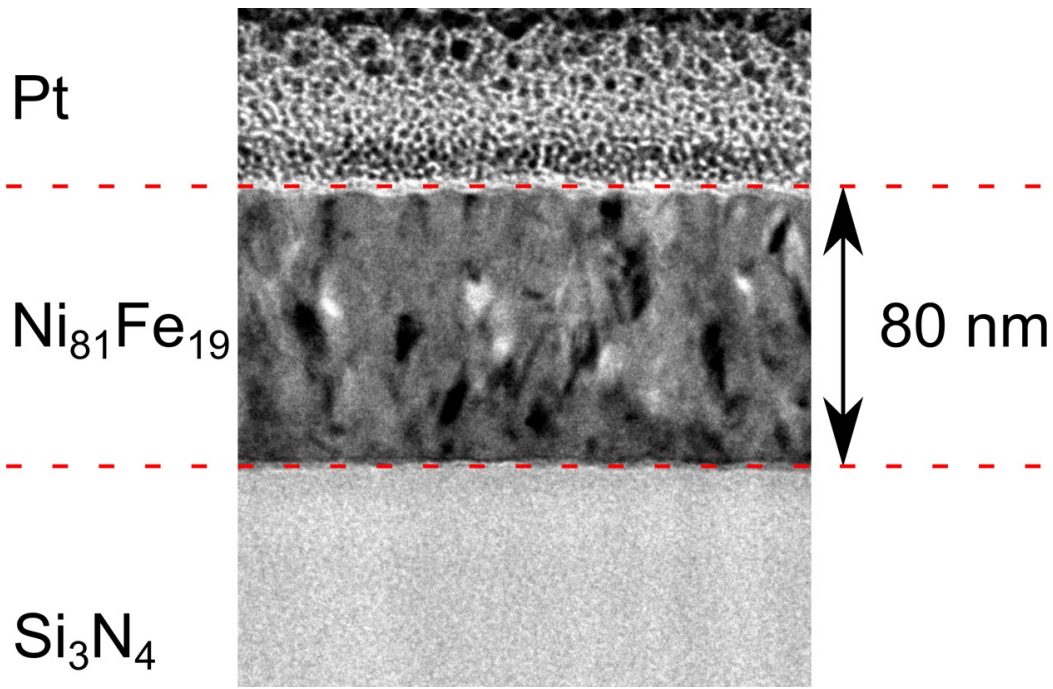

Figure S.10: Bright field transmission electron micrograph of a cross-sectional lamella extracted from a Permalloy reference sample.